\documentstyle [12pt,epsfig]{article} 
\makeatletter
\@addtoreset{equation}{section}
\makeatother

\newcommand{\ba}{\begin{eqnarray}}
\newcommand{\ea}{\end{eqnarray}}

\begin{document}
\newcommand{\BS}{\bigskip}
\newcommand{\SECTION}[1]{\BS{\large\section{\bf #1}}}
\newcommand{\SUBSECTION}[1]{\BS{\large\subsection{\bf #1}}}
\newcommand{\SUBSUBSECTION}[1]{\BS{\large\subsubsection{\bf #1}}}

\begin{titlepage}
\begin{center}
\vspace*{2cm}
{\large \bf Classical Electromagnetism as a Consequence of Coulomb's Law,
 Special Relativity and Hamilton's Principle and its Relationship to Quantum Electrodynamics
 \footnote{This paper is dedicated to the memory of Valentine Telegdi}}
\vspace*{1.5cm}
\end{center}
\begin{center}
{\bf J.H.Field }
\end{center}
\begin{center}
{ 
D\'{e}partement de Physique Nucl\'{e}aire et Corpusculaire
 Universit\'{e} de Gen\`{e}ve . 24, quai Ernest-Ansermet
 CH-1211 Gen\`{e}ve 4.}
\end{center}
\begin{center}
{e-mail; john.field@cern.ch}
\end{center}
\vspace*{2cm}
\begin{abstract}
  It is demonstrated how all the mechanical equations
  of Classical Electromagnetism (CEM) may be derived from only Coulomb's inverse square force law,
  special relativity and Hamilton's Principle. The instantaneous nature of the 
  Coulomb force in the centre-of-mass frame of two interacting charged objects,
  mediated by the exchange of space-like virtual photons, is predicted by QED.
  The interaction Lagrangian of QED is shown to be identical, in the appropriate
  limit, to the potential energy term in the Lorentz-invariant Lagrangian of CEM. 
  A comparison is made with the Feynman-Wheeler action-at-a-distance formulation 
  of CEM.
\end{abstract}
\vspace*{1cm}

{\it Keywords};  Special Relativity, Classical Electrodynamics.
\newline
\vspace*{1cm}
 PACS 03.30+p 03.50.De
\end{titlepage}

\SECTION{\bf{Introduction}}
   
    At the beginning of Book III of the Principia~\cite{Principia} Newton introduced
     four `Rules of Reasoning in Philosophy'. The first of them was:
   \par  {\tt We are to admit no more causes of natural things than such as are 
    both \newline true and sufficient to explain their appearences.}
    \par It is still a salutary exercise to apply this simple principle to any domain of 
    science. What are fundamental and truly important in the scientific description
    of phenomena are those concepts that cannot be discarded without destroying the
    predictive power of the theory. The most powerful, the best, scientific theory
    is that which describes the widest possible range of natural phenomena in terms
    of the minimum number of essential (i.e. non-discardable) concepts. It is the 
   aim of the present paper to apply this precept of Newton to Classical Electromagnetism
   (CEM). The relation of CEM to Quantum Electrodynamics (QED), in the attempt
   to obtain a deeper physical understanding of the former, will also be discussed.
   \par From the work of Coulomb, Amp\`{e}re and Faraday on, the basic phenomena
    of CEM, i.e. what are actually observed in experiments, are the forces between 
    electric charges at rest or in motion, or the dynamical consequences of such
    forces. The force between two static charges is given by Coulomb's inverse
    square law. This law will taken as a postulate in the following, but no other
    dynamical concept or theoretical construction will be introduced as an
    independent hypothesis in order to build up the theory. Later, it will be 
    seen that, in QED, this law is a necessary consequence of the existence of,
    and exchange of, space-like virtual photons between the electric charges.
    \par It will be assumed throughout that the system of interacting electric
    charges is a conservative one, in Classical Mechanics, and so may be 
    described by a Lagrangian that is a function of the coordinates and velocities
    of the charges, but does not depend explicitly on the time. Calculating
    the Action from the Lagrangian of the system and applying Hamilton's
    Principle that the Action be an extremum with respect to variation of
    the space-time trajectories of the charges, yields, in the well-known
    manner, the Lagrange equations that provide a complete dynamical description
    of the system~\cite{Golds1}. 
    \par It is further required that the physical description be consistent
    with Special Relativity. For this, the Lagrangian must be a
    Lorentz scalar. To introduce the method to be used to construct the
    Lagrangian, which is likely to be familiar only to particle physicists,
    I quote a passage taken from some lecture notes by R.Hagedorn~\cite{Hagedorn} on
    relativistic kinematics dating from some four decades ago:
     \par {\tt If a question is of such a nature that its answer will be always the 
        \newline same, no matter in which Lorentz system one starts, then it is 
      possible to formulate the answer entirely with the help of those invariants
      which one \newline can build with the available four vectors. One then finds the
      answer in a particular Lorentz system which one can choose freely and in 
      such a way \newline that the answer there is obvious and most easy. One looks then
     how the \newline invariants appear in this particular system, expresses the answer
    to the \newline problem by these invariants and one has found at the same time aleady the
    \newline general answer... It is worthwhile to devote some thinking to this method
    of calculation until one has completely understood that there is really 
    no jugglery or guesswork in it and that it is absolutely safe.}
     \par  It is important to stress the last sentence in this passage in relation
    to the word `true' in Newton's philosophical precept quoted above. Just the method
    outlined above was used to derive the Bargmann-Michel-Telegdi (BMT) equation for spin motion in  
    arbitary magnetic and electric fields~\cite{BMT}. 
    \par  It will be demonstrated in the following that it is sufficient to apply
    Hagedorn's programme to the simplest possible non-trivial
    electrodynamical system that may be considered: two mutually interacting 
    electric charges, in order to derive all the mechanical equations of CEM, as well
   as Maxwell's equations, with Coulomb's inverse square law as the only dynamical
   hypothesis. The 'mechanical' equations comprise the relativistic generalisation of the Biot and Savart Law,
        the  Lorentz force equation and those describing electromagnetic induction effects with 
    uniformly moving source currents  and test charges~\footnote{Not included are induction effects related
    to AC currents, where source charges are accelerated. Although described, in an identical manner, by the
     Faraday-Lenz Law, as non-accelerated charges, real as well as virtual photons must be taken
     into account, at the fundamental level, in this case. For uniformly moving charges, no real photons are created.}    
   \par An aspect that is not touched upon in the above programme is radiation.
     In this case a fundamental classical description of the phenomenon,
     in the sense of Newton's precept, is not possible and Quantum Mechanics
     must be invoked. In the language of QED, the existence of real photons
    as well as the virtual photons responsible for the
    Coulomb force, must be admitted. Indeed, extra degrees of freedom must
    be added to the Lagrangian to describe the propagation of real photons
    and their interaction with electric charges. Also the corresponding potentials and fields
    are retarded, not instantaneous. A brief comment is made in the
    concluding section on the relation of Maxwell's equations to radiation phenomena;
    however, no detailed comparison with QED is attempted. 
     \par It is also assumed throughout the paper that the effects of gravitation,
     that is of the curvature of space-time, on the interaction between the 
     charged physical objects considered, may be neglected.

 \SECTION{\bf{Lorentz Invariant Lagrangian for Two Mutually Interacting
   Electrically Charged Objects}}
   Two physical objects O$_1$ and O$_2$ of masses $m_1$ and $m_2$ and electric charges
    $q_1$ and $q_2$, respectively, are assumed to be in spatial proximity, far from 
    all other electric charges, so that they interact electromagnetically, but
     are subjected to no external forces. The 
    spatial positions of  O$_1$ and O$_2$ are specified, relative to their common
    center of energy,  by the vectors $\vec{r}_1$ and $\vec{r}_2$ respectively.
    The spatial distance separating the two objects in their common center-of-mass (CM)
    frame: $r_{12} = r_{21}$ is given by the modulus of the vectors $\vec{r}_{12}$,
    $\vec{r}_{21}$  where:
     \begin{equation}
    \vec{r}_{12} = - \vec{r}_{21} = \vec{r}_1-\vec{r}_2
    \end{equation}   
  \par The non-relativistic (NR) Lagrangian describing the motion of the objects O$_1$ and  O$_2$
     in their overall CM frame is~\cite{GoldsteinNRL}\footnote{Gaussian electromagnetic units are used.}
     \begin{equation}
      L_{NR}( \vec{r}_1,\vec{v}_1;\vec{r}_2,\vec{v}_2) \equiv T_1+T_2-V
       = \frac{1}{2}m_1 v_1^2+ \frac{1}{2}m_2 v_2^2 -\frac{q_1q_2}{r_{12}}
    \end{equation} 
  $T_1$, $v_1$ ($T_2$, $v_2$) are the kinetic energies and velocities, respectively of  O$_1$ (O$_2$) and $V$
       is the potential energy of the system.
   A Lorentz-invariant Lagrangian describing the system  O$_1$, O$_2$ will now be constructed in such a way
 that it reduces to Eqn(2.2) in the non-relativistic limit. The Lagrangian must be a Lorentz scalar
  constructed from the 4-vectors\footnote{ From translational invariance,
  the interaction between the objects does not depend upon the absolute positions of the
   objects, but only on their relative spatial separation: $|\vec{x}_1- \vec{x}_2|$. Therefore the
    dependence of the Lagrangian on the independent 4-vector $x_1+x_2$ may be neglected.}:
   $x_1-x_2$, $u_1$ and $u_2$ that completely specify the spatial
  and kinematical configuration of the interacting system. Here $\vec{x}_1 = \vec{r}_1$,
  $\vec{x}_2 = \vec{r}_2$ and the `4-vector velocity', $u$, is defined as:
    \begin{equation}
      u \equiv \frac{d x}{d \tau} = \gamma \frac{d~}{dt}(ct;\vec{x}) = (\gamma c ; \gamma \vec{v})
     \end{equation}
     where $\tau$ is the proper time of the object, $\gamma \equiv 1/\sqrt{1-\beta^2}$ and $\beta \equiv v/c$.
      In general, the Lagrangian may
    depend on the following six Lorentz invariants, constructed from the relevant 4-vectors:
     \[ (x_1-x_2)^2,~~~~ u_1 \cdot (x_1-x_2),~~~~  u_2 \cdot (x_1-x_2),~~~~ u_1^2,~~~~  u_2^2,~~~~ u_1 \cdot u_2  \]
     so that the Lagrangian may be written as:
    \begin{equation}
    L(x_1,u_1;x_2,u_2) =\alpha_0+ \alpha_1 (x_1-x_2)^2+ \alpha_2 u_1 \cdot (x_1-x_2) + \alpha_3 u_2 \cdot (x_1-x_2)
     +  \alpha_4 u_1^2  +  \alpha_5 u_2^2 + \alpha_6  u_1 \cdot u_2 
    \end{equation}
    where the coefficients $\alpha_0$-$\alpha_6$ are Lorentz-scalars that may also be, in general,
   arbitary functions of the six Lorentz invariants listed above.
    Taking the NR limit:
    \[ u_1 \rightarrow (c; \vec{v}_1),~~~ u_2 \rightarrow (c; \vec{v}_2)\]
   gives\footnote{Note that the term containing $\vec{v}_1 \cdot \vec{v}_2$ vanishes in the NR limit
      where terms of O( $\beta_1 \beta_2$) are neglected:
     $u_1 \cdot u_2 \rightarrow c^2(1-\vec{v}_1 \cdot \vec{v}_2/c^2)= c^2 +  O( \beta_1 \beta_2)$.}:
    \begin{equation}
    L(x_1,u_1;x_2,u_2) = \alpha_0 - \alpha_1 r_{12}^2 -  \alpha_2 \vec{v}_1 \cdot \vec{r}_{12}-
      \alpha_3 \vec{v}_2 \cdot \vec{r}_{12}
     - \alpha_4 v_1^2 - \alpha_5 v_2^2 +  ( \alpha_4 + \alpha_5 +\alpha_6)  c^2
    \end{equation}
      where a time-like metric is chosen for 4-vector products.
    Note that  $\vec{x}_1$ and $\vec{x}_2$ are defined at the same time, $t$,
    in the CM frame of
    O$_1$ and O$_2$, so that $t_1=t_2 = t$ in the 4-vectors $x_1$ and $x_2$. Thus the Coulomb interaction
    is assumed to be instantaneous in the CM frame. As discussed in Section 6 below, such behaviour is 
    a prediction of QED. 
    Consistency  between Eqns(2.2) and (2.5)  requires that\footnote{The symmetry of the Lagrangian
     with respect to the labels 1,2 requires that the term $\alpha_6  u_1 \cdot u_2$ be
    identified with the potential energy term in (2.2).}: 
     \begin{equation}
      \alpha_1 =  \alpha_2 = \alpha_3 = 0, ~~\alpha_4 = -\frac{m_1^2}{2}, ~~\alpha_5  = -\frac{m_2^2}{2},
       ~~ \alpha_6 = -\frac{q_1 q_2}{c^2 r_{12}}, ~~ \alpha_0 + (\alpha_4 + \alpha_5)c^2 = 0~~~ 
     \end{equation}
     The choice $\alpha_0 = c^2(m_1^2+m_2^2)/2$  satisfies the last condition in (2.6) and yields for
      the Lorentz-scalar Lagrangian:
        \begin{equation}
        L(x_1,u_1;x_2,u_2) = -\frac{m_1 u_1^2}{2} -\frac{m_2 u_2^2}{2} - \frac{j_1 \cdot j_2}{c^2 r_{12}}
      \end{equation} 
      Where the current 4-vectors: $j_1 \equiv q_1 u_1$ and $j_2 \equiv q_2 u_2$ have been introduced.
      This Lagrangian may be written in a manifestly Lorentz-invariant manner by noting that:
      \[ x_1-x_2 = (0;\vec{x}_1-\vec{x}_2) = (0; \vec{r}_{12}) \] 
       so that $r_{12} = \sqrt{-(x_1-x_2)^2}$ and
        \begin{equation}
        L(x_1,u_1;x_2,u_2) = -\frac{m_1 u_1^2}{2} -\frac{m_2 u_2^2}{2}
        - \frac{j_1 \cdot j_2}{c^2  \sqrt{-(x_1-x_2)^2}}
      \end{equation}
      The Lagrangian (2.7), when substituted into the covariant Lagrange equations derived from
      Hamilton's Principle~\cite{Golds1}:
       \begin{equation}
       \frac{d~}{d \tau}\left(\frac{\partial L}{\partial u_i^{\mu}}\right)
        -\frac{\partial L}{\partial x_i^{\mu}} = 0~~~(i=1,2;~\mu = 0,1,2,3)  :
        \end{equation}      
       is shown in the following Sections to enable all the
     concepts and equations of CEM concerning inter-charge forces, in the absence of radiation,
      to be derived without introducing any further postulate.
     Note that, since the Lagrangian (2.7) is a Lorentz scalar, it provides a description
    of the motion of O$_1$ and  O$_2$ in any inertial reference frame.
  
 \SECTION{\bf{The 4-vector Potential, Electric and Magnetic Fields, the Lorentz Force Equation
    and the Biot and Savart Law}}
  Considering only the motion of O$_1$, introducing the `4-vector potential', $A_2$, according
  to the definition:
  \begin{equation}
    A_2 \equiv \frac{j_2}{c r_{12}}
  \end{equation}
  the well-known~\cite{Golds2} Lorentz-invariant Lagrangian describing the motion of the
  object O$_1$ in the  `electromagnetic field created by the object O$_2$':
   \begin{equation}
        L(x_1,u_1) = -\frac{m_1 u_1^2}{2}  - \frac{1}{c} q_1 u_1 \cdot A_2
   \end{equation}
   is recovered. In the same way, the motion of  O$_2$ in the
    `electromagnetic field created by the object  O$_1$' is given by the invariant Lagrangian:
  \begin{equation}
        L(x_2,u_2) = -\frac{m_2 u_2^2}{2}  -  \frac{1}{c} q_2 u_2 \cdot A_1
   \end{equation}
  where:
    \begin{equation}
    A_1 \equiv \frac{j_1}{c r_{12}}
  \end{equation}
   To now introduce the concepts of distinct `electric' and `magnetic' fields it is 
   sufficient to consider only the motion of O$_1$. To simplify the equations the 
   labels `1' and `2' will be dropped in Eqn(3.2) and the following notation is used
    for spatial partial derivatives:
    \begin{equation}
    \partial_i = -\partial^i \equiv \frac{\partial~}{\partial x^i} \equiv \nabla_i~~~(i=1,2,3)
  \end{equation}
   The Lagrangian (3.2) is now introduced into the Lagrange equations (2.9). Considering the
   1 spatial components of the 4-vectors, the first term on the LHS of Eqn(2.9) is:
       \begin{equation}
       \frac{d~}{d \tau}\left(\frac{\partial L}{\partial u^1}\right) =  \frac{d~}{d \tau}
        (mu^1+\frac{q}{c} A^1) = \gamma(m \frac{d u^1}{dt}+\frac{q}{c}\frac{d A^1}{dt})
        \end{equation}
      and the second is:
         \begin{equation}
 -\frac{\partial L}{\partial x^1} = -\frac{q}{c} u \cdot(\partial^1 A)
  \end{equation}
    Combining Eqns(2.9), (3.6) and (3.7)and transposing:
       \begin{equation}
   \gamma m \frac{d u^1}{dt} =  \gamma  \frac{d p^1}{dt} = \frac{q}{c}[ u \cdot(\partial^1 A)
    -\gamma\frac{d A^1}{dt}]
       \end{equation}
    where the `energy-momentum 4-vector' $p \equiv m u$ has been introduced. Substituting the
    Euler formula for the total time derivative\footnote{The implict time dependence of $A^1$ in the first
     term on the right side of (3.9) arises from the instantaneous motion of the `source' O$_2$,
   whereas the remaining terms
      describe the variation of $A^1$ due to the motion of O$_1$.}:
      \begin{equation}
   \frac{d A^1}{dt} =  \frac{\partial A^1}{\partial t} - v^1 \partial^1 A^1 - v^2 \partial^2 A^1
    - v^3 \partial^3 A^1 
          \end{equation}
    into (3.8), writing out explicitly the 4-vector product $ u \cdot(\partial^1 A)$, and cancelling a common 
    factor $\gamma$ from each term, gives:
   \begin{equation}
     \frac{d p^1}{dt} = \frac{q}{c}\left[ c \partial^1 A^0- \frac{\partial A^1}{\partial t}
      +v^2(\partial^2 A^1- \partial^1 A^2) -v^3(\partial^1 A^3- \partial^3 A^1) \right] 
        \end{equation}
      Introducing now 3-vector `electric' and `magnetic' fields, $E^i$ and $B^i$ respectively,
      according to the definitions:
      \begin{equation}
  E^i \equiv  \partial^i A^0- \frac{1}{c}\frac{\partial A^i}{\partial t} = \partial^i A^0- \partial^0 A^i  
      \end{equation} 
     and 
       \begin{equation}
  B^k \equiv -\epsilon_{ijk}(\partial^i A^j- \partial^j A^i) = (\vec{\nabla} \times \vec{A})^k
     \end{equation}
     where $\epsilon_{ijk}$ is the alternating tensor equal to $+1(-1)$ when $ijk$ is an
    even (odd) permutation of 123, and zero otherwise, 
    enables Eqn(3.10) to be written as the compact expression:
       \begin{equation}
     \frac{d p^1}{dt} = q \left[ E^1 + \frac{1}{c}(\vec{v} \times \vec{B})^1 \right]
      \end{equation}
      which is the 1 component of the Lorentz force equation. The 2 and 3 components 
    are derived by cyclic permutations of the indices 1,2,3 in Eqn(3.10), yielding finally the
    3-vector Lorentz force equation:
      \begin{equation}
     \frac{d\vec{p}}{dt} = q \left[\vec{E} + \frac{\vec{v}}{c} \times \vec{B} \right]
      \end{equation}
   The concepts of `electric' and `magnetic' fields have therefore appeared naturally 
  as a means to simplify the Lorentz force equation (3.10). However, the RHS of this equation 
 is completely defined, via Eqn(3.1), by the 4-vector current $j_2$, the spatial separation $r_{12}$
 of O$_1$ and O$_2$ and the 3-velocity of O$_1$, so that the 4-vector potential $A$ may be 
   eliminated from the Lorentz force equation. Substituting the definition of $A$ from Eqn(3.1)
   into Eqns(3.11) and (3.12), and restoring the labels of quantities associated with O$_2$,
   gives\footnote{Note that the partial time derivative in (3.11) implies
   that $\vec{x}_1$ but not $\vec{x}_2$ is held constant. The implicit time variation of $A^1$
   in (3.11) then has contributions from both $\vec{j_2}$ and  $\vec{x}_2$ which yield,
    respectively, the last two terms on the right side of (3.15).}:
     \begin{equation}
      \vec{E} = \frac{j_2^0 \vec{r}}{c r^3} -\frac{1}{c^2 r}\frac{d \vec{j_2}}{d t}
    -\frac{\vec{j_2}}{c^2}\frac{(\vec{r} \cdot \vec{v_2})}{r^3}
     \end{equation}
    \begin{equation}
     \vec{B} = \frac{q_2 \gamma_2 (\vec{v_2} \times \vec{r})}{c r^3} = \frac{\vec{j_2} \times  \vec{r}}{c r^3}
   \end{equation} 
    where $\vec{r} \equiv \vec{r}_{12}$. 
   Eqn(3.16) is the relativistic generalisation of the Biot and Savart Law. It differs
     from the usual CEM formula by a factor $\gamma_2$. Note that the electric field is,
    in general, non-radial. The non-radial part of the field, associated with the last term on the
   right side of (3.15), originates in the second term on the right side of (3.11). This is the
   electric field that is associated with the time variation of the magnetic field in the
   Faraday-Lenz Law. For the case of a source charge in uniform motion in the $x$-direction,
   with velocity $v_2$, the electric and magnetic fields given by (3.15) and (3.16) at the field
    point $\vec{r} = \hat{\imath} \cos \psi + \hat{\jmath} \sin \psi$ are:
     \begin{eqnarray}
       \vec{E} & = & \frac{q}{r^2}\left[\frac{\hat{\imath} \cos \psi}{\gamma_2}+ \gamma_2  \hat{\jmath} \sin \psi \right] \\
 \vec{B} & = & \frac{\vec{v}_2 \times \vec{E}}{c}
      \end{eqnarray}
     where $\hat{\imath}$ and $\hat{\jmath}$ are unit vectors in the $x$- and $y$-directions. These equations
     may be compared with the pre-relativistic Heaviside~\cite{Heaviside} formulae for this case:
     \begin{eqnarray}
       \vec{E}(H) & = & \frac{q \vec{r}}{r^3 \gamma_2^2(1- \beta_2^2 \sin^2 \psi)^{\frac{3}{2}}} \\
 \vec{B}(H) & = & \frac{\vec{v}_2 \times \vec{E}(H)}{c}
      \end{eqnarray}
     The fields $\vec{E}(H)$ and  $\vec{B}(H)$ are also the `present time' fields as derived~\cite{PP} from the
     retarded Li\'{e}nard-Wiechert potentials~\cite{LW}. By considering a simple two-charge `magnet', in a 
    particular spatial configuration, either
     in motion or at rest, it has been shown~\cite{JHFEMI} that the radial electric field of (3.19) predicts a vanishing
     induction effect for a moving magnet and stationary test charge. In the same configuration
     (3.17) predicts the same induction force on the test charge as the Faraday-Lenz Law. The Heaviside
     formulae are therefore valid only to first order in $\beta$, in which case the predictions
      of (3.19) and (3.20) are the same as those of (3.17) and (3.18). It is interesting to recall that
      just this problem, of induction in different frames of reference, was discussed in the Introduction
       of Einstein's 1905 special relativity paper~\cite{Ein1}. 
  \par Substitution of (3.15) and 
   (3.16) into (3.14) and restoring the labels associated with O$_1$  yields the `fieldless'
    Lorentz force equations\footnote{The right sides of these equations 
   are `forces' according to the relativistic generalisation of
    Newton's Second Law. In fact, however, the force concept does not appear at
    any place in their derivation. Also the relativistic 3-momentum $\vec{p} = \gamma \vec{\beta}m c$
    appears naturally in the equations as a necessary consequence of the initial postulates.
     For an interesting recent discussion 
    of the force concept in modern physics see~\cite{Wilczek}.} for two, discrete, mutually electromagnetically interacting, physical objects:
    \begin{eqnarray} 
      \frac{d\vec{p_1}}{dt} &  = & \frac{q_1}{c}\left[\frac{ j_2^0\vec{r} +  \vec{\beta}_1 \times
     (\vec{j_2} \times \vec{r})}{r^3} -\frac{1}{c r}\frac{d \vec{j_2}}{d t}-\vec{j_2}
     \frac{(\vec{r} \cdot \vec{\beta}_2)}{r^3} 
      \right] \\
       \frac{d\vec{p_2}}{dt}  &  = & -\frac{q_2}{c}\left[\frac{ j_1^0\vec{r} +\vec{\beta}_2 \times 
  (\vec{j_1} \times \vec{r})}{r^3}+\frac{1}{c r}\frac{d \vec{j_1
}}{d t} -\vec{j_1}
     \frac{(\vec{r} \cdot \vec{\beta}_1)}{r^3}
    \right] 
     \end{eqnarray} 
     It may be thought that the terms $\simeq 1/r$ should be assocated with radiative procesees
 (see Section 7 below) but they are in fact of particle-kinetic nature. Since $\vec{j} = (q/m)\vec{p}$
  the two differential equations are coupled via the $d\vec{j}/dt$ terms on the right sides of each.
   The solution of these equations for the case of circular Keplerian orbits has been derived~\cite{JHFRSKO}.
   One result obtained is the relativistiic generalisation of Kepler's Third Law of planetary motion
     for this case:

     \begin{equation} 
     \tau^2 =  \frac{(2 \pi)^2 {\cal E}^* \left[1 -\frac{(q_1 q_2)^2}{m_1 m_2 c^4  r^2}\right] r^3}
   {|q_1||q_2|(1+\beta_1 \beta_2)}
   \end{equation}
 where
   \begin{equation}
    {\cal E}^* \equiv \frac{{\cal E}_1^*{\cal E}_2^*}{{\cal E}_1^*+{\cal E}_2^*}  
     \end{equation}
  and
           \begin{eqnarray} 
 {\cal E}_1^* & \equiv & \frac{\gamma_1 m_1 c^2}{\gamma_2-\frac{|q_1||q_2| \gamma_1}{m_2 c^2 r}} \\
  {\cal E}_2^* & \equiv & \frac{\gamma_2 m_2 c^2}{\gamma_1-\frac{|q_1||q_2| \gamma_2}{m_1 c^2 r}} 
      \end{eqnarray} 

    Eqn(3.23) gives the period, $\tau$, of two objects of mass $m_1$ and $m_2$ with (opposite)
    electric charges $q_1$ and $q_2$, in circular orbits around their common center of energy,
    separated by the distance $r$. The $d \vec{j}/d t$ terms in (3.21) and (3.22) give the
      terms $\simeq 1/r$ in the denominators on the right sides of (3.25) and (3.26).
     These terms effectively modify the masses of the objects due to the electromagnetic
      interaction.
   \par It is also demonstrated in Ref.\cite{JHFRSKO} that stable, circular, Keplerian
    orbits are impossible under the retarded forces generated by Li\'{e}nard-Wiechert potentials.

    \par Considering now the time components of the 4-vectors in (2.11), the first term on the LHS
  is: 
       \begin{equation}
       \frac{d~}{d \tau}\left(\frac{\partial L}{\partial u^0}\right) = 
        \gamma(-m \frac{d u^0}{dt}-\frac{q}{c}\frac{d A^0}{dt})
        \end{equation}
      while the second is:
         \begin{equation}
 -\frac{\partial L}{\partial x^0} = \frac{q}{c} u \cdot(\partial^0 A)
    = \frac{q}{c} u \cdot \left(\frac{1}{c}\frac{\partial A}{\partial t}\right)
  \end{equation}
    
 Substituting (3.27) and (3.28) into (2.9) and rearranging gives:
         \begin{equation}
    \gamma \frac{d {\cal E}}{d t}=\frac{q}{c}\left[ u \cdot \left(\frac{1}{c}\frac{\partial A}{\partial t}\right)
     -\gamma\frac{q}{c}\frac{d A^0}{dt}\right]
        \end{equation}
   where $ {\cal E} \equiv m u^0 c$ is the relativistic energy of O$_1$. Using the Euler formula
    (3.9) to express $d A_0/d t$ in terms of partial derivatives, and writing out the different
     terms in the 4-vector scalar products, the terms $\partial A^0/ \partial t$ are seen to
     cancel. Dividing out the factor $\gamma$ on both sides of the equation then gives the 
     result:
   \begin{equation}
     \frac{d {\cal E}}{d t} = q[v_1(\partial^1 A^0-\partial^0 A^1)+v_2(\partial^2 A^0-\partial^0 A^2)
     +v_3(\partial^3 A^0-\partial^0 A^3)] = q \vec{v} \cdot \vec{E}
         \end{equation}
     where $\vec{E}$ is the electric field defined in (3.11). Restoring now the labels of  O$_1$ and
      O$_2$ gives the `fieldless' equations for the time derivatives of their relativistic energies:

    \begin{eqnarray} 
      \frac{d{\cal E}_1}{dt} &  = & q_1 \left[ j_2^0 \frac{\vec{\beta_1} \cdot \vec{r}}{r^3}
    -\frac{1}{c r} \vec{\beta_1} \cdot \frac{d \vec{j_2}}{d t}-
     \frac{(\vec{\beta_1} \cdot \vec{j_2})(\vec{r} \cdot \vec{\beta_2})}{r^3} 
      \right] \\
     \frac{d{\cal E}_2}{dt} &  = & -q_2 \left[ j_1^0  \frac{\vec{\beta_2} \cdot \vec{r}}{r^3}
    +\frac{1}{c r} \vec{\beta_2} \cdot \frac{d \vec{j_1}}{d t}+
     \frac{(\vec{\beta_2} \cdot \vec{j_1})(\vec{r} \cdot \vec{\beta_1})}{r^3} 
      \right]
     \end{eqnarray}  
      
    The equations (3.21),(3.22) and (3.31),(3.32) give a complete description of
    the purely mechanical aspects of CEM (that is, neglecting radiative effects) for two massive, electrically
    charged, objects 
    interacting mutually through electromagnetic forces.
     \par The Lagrangian (2.7) is readily generalised to describe the mutual electromagnetic interactions of
     an arbitary number of charged objects:
        \begin{equation}
        L(x_1,u_1;x_2,u_2; ...,x_n,u_n) = -\frac{1}{2} \sum_{i=1}^{n} m_i u_i^2 -
         \frac{1}{c^2} \sum_{i>j}q_i q_j \frac{u_i \cdot u_j}{r_{ij}}
      \end{equation} 
     Here $r_{ij} = |\vec{r}_i-\vec{r}_j|$ where $\vec{r}_i$ and $\vec{r}_j$ specify the positions
     of O$_i$ and  O$_j$, respectively, relative to the centre-of-energy on the $n$ interacting objects.
      Note that, as all these distances are specified at a fixed time in the overall CM frame
   of the objects, the  $r_{ij}$ are Lorentz invariant quantities, similar to $r_{12}$ in Eqn(2.7).
   See also~\cite{JHF1} for a general discussion of such invariant length intervals. The Lagrangian 
    describing the motion of the object $i$ `in the electromagnetic field of' the remaining $n-1$
    objects may be derived from Eqn(3.33):
   \begin{equation}
        L(x_i,u_i) = -\frac{m_i u_i^2}{2}  - \frac{1}{c} q_i  u_i \cdot A(n-1)
   \end{equation} 
    where 
     \begin{equation}
      A(n-1) \equiv \sum_{j \ne i}^{n}\frac{q_j u_j}{r_{ij}}
     =  \sum_{j \ne i}^{n}\frac{j_j}{r_{ij}}
    \end{equation}  
   This equation embodies the classical superposition principle for the electromagnetic
   4-vector potential, and hence, via the linear equations (3.11) and (3.12),
   that for the electric and magnetic fields.

 \SECTION{\bf{Derivation of Maxwell's Equations}}
  Writing out explicitly the spatial components of the quantity $\vec{\nabla} \cdot \vec{B}$ using
  the definition of $\vec{B}$, Eqn(3.12):
  \begin{eqnarray}
  \partial^1 B^1 & = & \partial^1 \partial^3 A^2 -  \partial^1 \partial^2 A^3 \\
  \partial^2 B^2 & = & \partial^2 \partial^1 A^3 -  \partial^2 \partial^3 A^1 \\
   \partial^3 B^3 & = & \partial^3 \partial^2 A^1 -  \partial^3 \partial^1 A^2
  \end{eqnarray}
   it follows, since $\partial^i\partial^j=\partial^j\partial^i~~(i,j=1,2,3)$ that, on summing
   Eqns(4.1), (4.2) and (4.3),
   \begin{equation}
    \vec{\nabla} \cdot \vec{B} = -(\partial^1 B^1+ \partial^2 B^2+  \partial^3 B^3) = 0
    \end{equation} 
    which is the magnetostatic Maxwell equation. Since $\vec{B} \equiv \vec{\nabla} \times \vec{A}$, (4.4) can also
     be seen to follow from the 3-vector identity $\vec{a} \cdot (\vec{a} \times \vec{b}) \equiv 0$ for arbitary
     $\vec{a}$ and $\vec{b}$.   
    \par The Faraday-Lenz Law follows directly from the defining equations Eqn(3.11), (3.12)
   of the electric and magnetic fields. Taking the curl of both sides of the 3-vector form 
   of Eqn(3.11) with $\vec{\nabla}$ gives:
    \begin{equation}
     \vec{\nabla} \times \vec{E} =-\vec{\nabla} \times (\vec{\nabla} A^0)
    - \frac{\partial~}{\partial t}(\vec{\nabla} \times \vec{A})    
  \end{equation}
    Since $\vec{\nabla} \times (\vec{\nabla} \phi) = {\rm curl} ({\rm div} \phi) = 0$  for an arbitary scalar $\phi$,
    the first term on the RHS of Eqn(4.5) vanishes. Subsituting the 3-vector
   form of Eqn(3.12) in the second term on the RHS of Eqn(4.5) then yields the 
    Faraday-Lenz Law:
  \begin{equation}
     \vec{\nabla} \times \vec{E} =   - \frac{1}{c} \frac{\partial \vec{B}}{\partial t}    
  \end{equation}
 \par The electrostatic Maxwell equation:
   \begin{equation}
    \vec{\nabla} \cdot \vec{E} = 4 \pi J^0
  \end{equation}
  is a well-known consequence of the inverse square law for a `static' electric field
  defined by only the first term on the RHS of Eqn(3.11) and Gauss' theorem~\cite{Jackson1}.
  The 4-vector current density: $J \equiv (c\rho; \vec{J})$, the 0 component of which appears in
  Eqn(4.7), is related to the currents, $j_i$, of elementary charges $q_i$ by the relation:
   \begin{equation}
   J = \frac{1}{V_R}\sum_{i \subset R} j_i
  \end{equation}
  where $V_R$ is the volume of a spatial region $R$. Hence $\rho = J^0/c$ is, in the
  non-relativistic limit where $\gamma \simeq 1$, the average 
  spatial density of electric charge in the region $R$. Conservation of electric charge
  requires that:
   \begin{equation}
     \frac{\partial \rho}{\partial t} +\vec{\nabla} \cdot \vec{J} = 0
   \end{equation}
   This continuity equation may be simply derived from the properties of the 
   4-vector product:
    \begin{equation}
  \partial \cdot j_i =  \partial^0 j_i^0 -\sum_{k=1}^3\partial^k j_i^k = q_i \left[
  c\frac{\partial \gamma_i}{\partial t}+\vec{\nabla} \cdot ( \gamma_i \vec{v_i}) \right]
   \end{equation}
   In the rest frame of the object
   O$_1$, $ \gamma_i-1 = | \vec{v_i}| = 0$, so that $ \partial \cdot j_i = 0$. Since
    $ \partial \cdot j_i$ is a Lorentz invariant this quantity then vanishes in all
   inertial refererence frames.  
   Taking the scalar product of $\partial$ and $J$ gives:
    \begin{equation}
      \partial \cdot J =  \frac{\partial \rho}{\partial t} +\vec{\nabla} \cdot \vec{J}
        = \frac{1}{V_R}\sum_{i \subset R} \partial \cdot j_i  = 0 
   \end{equation}
        Which is just the continuity equation (4.9).
    It can be seen that the conservation of electric charge is a consequence of its
   Lorentz-scalar nature, i.e. the charge $q_i$ in Eqn(4.10) does not depend on the
    frame in which $\vec{v_i}$ is evaluated. Indeed, the definition $j_i \equiv q_i  u_i$
    implies that $j_i \cdot j_i = q_i^2 u_i  \cdot  u_i =  c^2 q_i^2$, so that $q_i^2$
     is manifestly Lorentz invariant,
   in precise analogy with the mass of an object: $p_i \cdot p_i =  m_i^2 u_i  \cdot  u_i =  c^2 m_i^2$.
    Both $j_i$ and $p_i$ are proportional to the 4-vector velocity $u_i$.
 
    \par A relation similar to (4.9) is:
    \begin{equation}
    \frac{1}{c}\frac{\partial A^0}{\partial t}+\vec{\nabla} \cdot \vec{A}   = 0
   \end{equation}
  the so-called `Lorenz Condition'\footnote{ Not `Lorentz Condition', as found in many text
   books. See Reference~\cite{JackOkun}.}, which may also be written more simply as $\partial \cdot A = 0$.
    This relation is, in the present approach, not, as in conventional discussions of CEM,
    the result of a particular choice of gauge in the definition
    of $\vec{A}$, but an identity following from the definition of $A$ in Eqn(3.1). In fact as is easily
    shown:
     \begin{equation}
 \vec{\nabla} \cdot \vec{A} = -\frac{\vec{j} \cdot \vec{r}}{c r^3} = -\frac{1}{c}\frac{\partial A^0}{\partial t}
   \end{equation}
    Here the derivatives in $\vec{\nabla}$ are with respect to the `field point' $\vec{x}_1$ in contrast with 
    those in $\vec{\nabla}$  in Eqns(4.9)-(4.11), which are with respect to the spatial coordinate
  $\vec{x}_2$ of the object O$_2$ associated with the current $\vec{j}$. The partial time derivative
    in (4.12) is defined for $\vec{x}_1$ constant. The time variation of $A^0$ is then due solely
    to the time dependence of $\vec{x}_2$, which leads to the second member of (4.13). Eqn(4.12) shows that the 
   4-vector potential, like the current and energy-momentum 4-vectors corresponds to a conserved
   (Lorentz invariant) quantity: $A \cdot A = q^2 /r^2$ \footnote{$r$ is the manifestly
   invariant quantity $\sqrt{-(x_1-x_2)^2}$ that appears in eqn(2.10) above.}.
    So both $j$ and $A$ differ only by
   Lorentz invariant multiplicative factors from $p$ and $u$:
     \begin{equation} 
    c^2 = u \cdot u = \frac{p \cdot p }{m^2} = \frac{j \cdot j}{q^2} = c^2 r^2 \frac{A \cdot A}{q^2} 
    \end{equation}
     The relation (4.12) is found to be important in an interpretation of the
    electrodynmamic Maxwell equation,
  (4.20) below, as a description of radiation phenomena (creation of real photons). This point will be briefly
   discussed in Section 7. 

     \par The electrodynamic Maxwell equation (Amp\`{e}re's Law, including Maxwell's `displacement
     current') is derived immediately on writing the electrostatic Maxwell equation
    (4.7) in a covariant form. The latter then appears as an equation for the 0 component of 
     a 4-vector. The corresponding spatial components, written down simply by inspection,
     are Amp\`{e}re's Law.
      Writing Eqn(4.7) in 4-vector notation, and introducing also the `non-static' component
    of the electric field, given by the second term on the RHS of Eqn(3.11), gives:
    \begin{equation}
      (\sum_{i=1}^3 -\partial^i \partial^i) A^0 - \partial^0(\sum_{i=1}^3 -\partial^i A^i) = 4 \pi J^0
   \end{equation}
    Adding to Eqn(4.15) the identity:
     \[\partial^0 \partial^0 A^0 -\partial^0 \partial^0 A^0 = 0 \]
     gives:
     \begin{equation}
    (\partial \cdot \partial) A^0 -\partial^0(\partial \cdot A) =  4 \pi J^0
      \end{equation}
     Since the coefficients of $A^0$ and $-\partial^0$ are Lorentz scalars, the corresponding
     $i$th spatial component of the 4-vector $J$, is from the manifest covariance of Eqn(4.16),
     given by the equation:
         \begin{equation}
    (\partial \cdot \partial) A^i -\partial^i(\partial \cdot A) =  4 \pi J^i
     \end{equation}
     This is Amp\`{e}re's Law in 4-vector notation. In order to recover the more familiar 3-vector
     equation, the 4-vector potential must be eliminated in favour of the electric and 
    magnetic fields defined in Eqns(3.11) and (3.12) respectively. To do this, consider
    the contribution of the spatial parts (SP) of the 4-vector products on the LHS of Eqn(4.17) 
    to $J^1$. This gives:
   \begin{eqnarray} 
    4 \pi J^1(SP) & = & - \sum_{i=1}^3(\partial^i)^2 A^1+\partial^1 \sum_{i=1}^3(\partial^i A^i)
                    =   \sum_{i=1}^3\partial^i(\partial^1 A^i-\partial^i A^1)  \nonumber \\
          & = & (\partial^1)^2A^1 +\partial^2 \partial^1 A^2+ \partial^3 \partial^1 A^3
        - (\partial^1)^2A^1 -(\partial^2)^2A^1 -(\partial^3)^2A^1 \nonumber \\
           & = &  = -\partial^2(\partial^2 A^1 -\partial^1 A^2) + \partial^3(\partial^1 A^3 -\partial^3 A^1)     
         \nonumber \\
           & = & - \partial^2 B^3 +  \partial^3 B^2 = (\vec{\nabla} \times \vec{B})^1
    \end{eqnarray}
    where, in the fourth line the definition, Eqn(3.12), of the magnetic field has been used.
    The contribution of the temporal parts (TP) of the 4-vector products on the LHS of  Eqn(4.17) 
    to $J^1$ is:
 \begin{equation} 
    4 \pi J^1(TP) = (\partial^0)^2 A^1-\partial^1 \partial^0 A^0 = \partial^0 (\partial^0 A^1
   -\partial^1 A^0) = -\frac{1}{c} \frac{\partial E^1}{\partial t}       
  \end{equation}
   Adding the spatial and temporal contributions to $J^1$ from Eqns(4.18) and (4.19) gives the 1
   component of the electrodynamic Maxwell equation:
   \begin{equation}
     \vec{\nabla} \times \vec{B} -\frac{1}{c} \frac{\partial \vec{E}}{\partial t}  = 4 \pi \vec{J}
   \end{equation}
    The 2 and 3 components are obtained by cyclic permutation of the indices 1,2,3 in \newline
     Eqns(4.18) and (4.19). This derivation of  Amp\`{e}re's Law, starting from the electrostatic Maxwell
    equation, (4.7) has been previously given by Schwartz~\cite{Schwartz}, and, independently, by
    the present author in Reference~\cite{JHF2}, where it was noted that Eqn(4.17) may be derived
    from Eqn(4.16) using space-time exchange symmetry invariance.

  \SECTION{\bf{Fundamental Concepts and Different Levels of Mathematical Abstraction}}

   Equations (3.21),(3.22),(3.31) and (3.32) show that the dynamics of any system of mutually
    interacting electrically charged
    objects is completely specified by their masses, electric charges and 4-vector positions
    and velocities. Other useful and important concepts of CEM such as the 4-vector potential
 and electric and magnetic fields are completely specified, in terms of the geometrical and 
    kinematical configuration of the charged objects by Eqn(3.1) for $A^{\mu}$, Eqns(3.1)
     and (3.11) for $\vec{E}$ and 
    Eqns(3.1) and (3.12) for $\vec{B}$. Historically, of course, Faraday arrived at the concepts 
   of electric and magnetic fields in complete ignorance of the existence of elementary 
   electric charges or of Special Relativity. With our present-day understanding of both the existence
   of the former and the necessary constraints provided by the latter, it can be seen that both the
   4-vector potential and  electric and magnetic fields are, in fact, only convenient mathematical 
   abstractions. The 4-vector potential is at a first level of abstraction. The phenomenologically
   most useful concepts of CEM, the electric and magnetic fields are, in turn, completely
   specified by  $A^{\mu}$ and so are at a second level of abstraction from the fundamental
   and irreducible concepts (charged, interacting, physical objects) of the theory.
  \par Indeed, there is yet a third level of abstraction, the tensor $F^{\mu \nu}$ of the
   electromagnetic field defined as:
   \begin{equation}
   F^{\mu \nu} = \partial^{\mu} A^{\nu} - \partial^{\nu} A^{\mu}
   \end{equation}
   This description was introduced by Einstein in his original paper on General Relativity~\cite{Einstein1}
   in analogy with the tensor $G^{\mu \nu}$ of the classical gravitational field. It has the merit 
   of enabling the electrostatic and electrodynamic Maxwell equations to be written as a single compact
   equation\footnote{ The covariant operator $\partial_{\nu}$ is introduced by multiplying the
   contravariant operator $\partial^{\mu}$ by the metric tensor:
    $\partial_{\nu} = g_{\nu \mu} \partial^{\mu}$ where $ g_{\nu \mu}= 0$ for $\nu \ne \mu$ and
     $ g_{\mu \mu} =(1,-1,-1,-1)$. Repeated upper and lower indices: $\nu$,$\mu$ are summed
    over 0,1,2,3.}:
      \begin{equation}
      \partial_{\nu} F^{\mu \nu} = 4 \pi J^{\mu}
      \end{equation}
    As in the case of the introduction of electric and magnetic fields into the covariant Lorentz
    force equation (3.10) to obtain the 3-vector version (3.14), a cumbersome equation is reduced
   to an elegant one, at the the cost of introducing a higher level of mathematical abstraction.
    Viewed, however in the light of the strict criteria of Newton's precept,  $A^{\mu}$, $\vec{E}$,
    $\vec{B}$ and  $F^{\mu \nu}$, (although in the case of $\vec{E}$ and  $\vec{B}$ extremely
     useful phenomenologically) are certainly not `sufficient' to explain, in any fundamental 
     manner, the phenomena of CEM. On the contrary, as shown above, Coulomb's law and Special
      Relativity, given, of course, the {\it a priori} existence of charged physical objects,
     do provide such a fundamental description, in which the `fields' of electromagnetism
     appear naturally by mathematical substitution. If all that was known of CEM was Eqn(5.2),
     it is hard to see any logical path to derive from it the Lorentz Force, Biot and Savart
     and Faraday-Lenz Laws that actually describe the results of laboratory experiments in CEM.
      However these laws, Eqn(5.2) and the magnetostatic Maxwell equation
     (4.4) are all necessary
     consequences of Coulomb's Law, Special Relativity and Hamilton's Principle. The higher
     the level of mathematical abstraction, the more elegant the electrodynamic formulae
     appear to be, but the further removed they become from the physical realities of
     the subject.
     \par Although Einstein spent some decades of his life in the unsuccessful attempt
    to realise a unifying synthesis between the classical field tensors 
    $F^{\mu \nu}$ and $G^{\mu \nu}$ he still made a clear distinction between physical
    reality and mathematical abstraction\cite{Einstein2}:
  \par  {\tt We have seen, indeed, that in a more
    complete analysis the energy tensor\newline can be regarded only as a provisional means of representing
     matter. In \newline reality, matter consists of electrically charged particles, and is to be \newline
     regarded itself as a part, in fact the principle part, of the \newline electromagnetic field.}
   \par In fact, electrically charged particles and real and virtual photons (which are also particles)
    are the true irreducible concepts of CEM. These are not the `principle part' of the 
    electromagnetic field, but rather {\it replace it}
    in the most fundamental description of the phenomena of CEM.   
     \par Since the only dynamical postulate in CEM is Coulomb's Law, the only way to obtain
     a deeper physical understanding is by a deeper understanding of this Law. 
      Indeed, as will be discussed in the following Section, this does seem to be possible by considering
     the particle aspects of the  microscopic underlying QED process, which is basically M\o ller scattering:
     $e^-e^- \rightarrow e^-e^-$.

  \SECTION{\bf{Quantum Electrodynamical Foundations of Classical Electromagnetism}}
     If the electrodynamical force is transmitted by particle exchange, and it is assumed that the
    magnitude of the force is proportional to the number of interacting particles, which are emitted
    isotropically by the source, the inverse square law follows from spatial geometry and 
    conservation of the number of particles\footnote{A similar physical reasoning was followed  
    by  Kepler in his attempts to understand the gravitational force. As, however, the agents of
   force were constrained to propagate in the plane of a planetary orbit, rather than in 
   three spatial dimensions, a $1/r$ force law was predicted~\cite{Kepler}. Also, as a consequence
   of Kepler's Aristotelian understanding of dynamics, the force was conjectured to sweep
   the planets around the Sun in the transverse direction, rather than diverting them
   radially from their natural rectilinear motion, as in Newtonian dynamics.}.
    However, in the Coulomb interaction the exchanged
     particle is a virtual, not a real, photon. This means that it cannot always be considered to move in
    a particular direction in space-time. It will be shown below, however, that the Fourier transform
    of the momentum-space virtual photon propagator does yield a space-time propagator with the
    $1/r$ dependence of the Coulomb potential, which corresponds, in the classical limit, 
    to an inverse square force law. It is also shown that, in the CM frame of the interacting charged particles, this 
    interaction is instantaneous, as assumed in the derivation of the classical Lagrangian (2.7). 

     According to QED, the Biot and Savart and
     Lorentz Force Laws are the classical limit of M\o ller scattering for very large numbers of
     electrons at very large spatial separations. Conversely, M\o ller scattering is the
     quantum limit of the Biot and Savart and Lorentz Force Laws when each current contains
     a single electron and the spatial separation of the currents is very small. The fundamental
     quantum mechanical laws governing M\o ller scattering do not change when many electrons, 
     with macroscopic spatial separations, participate in the observed physical phenomenon.
     A more fundamental understanding of CEM is therefore provided, not by any kind of
     field concept, but by properly taking into account the existence of virtual
     photons, just as an analysis in terms of real photon production
     is mandatory for a fundamental description of the radiative processes
     of CEM, a subject beyond the scope of the present paper.
     \par The invariant QED  amplitude for M\o ller scattering by the exchange of 
     a single virtual photon\footnote{Actually there are two such amplitudes
     related by exchange of the identical final state electrons. In the
   present case, where the classical limit of CEM is under discussion,
  it suffices to consider only the amplitude given by (6.1) in the 
 limit $q^2 \rightarrow 0$. The contribution of the 
   second amplitude is negligible in this limit.} is given by the expression~\cite{HM1}
      \footnote{Here units with $\hbar = c = 1$ are assumed.}:
    \begin{equation} 
       T_{fi} =  -i \int \frac{{\cal J}^A(x_A) \cdot {\cal J}^B(x_A)}{q^2} d^4x_A
    \end{equation}
      The corresponding Feynman and momentum-space diagrams are shown in Fig.1. 
      The virtual photon is exchanged between the 4-vector currents ${\cal J}^A$ 
      and  ${\cal J}^B$ defined in terms of plane-wave solutions, $u_i$,~$u_f$ of the Dirac 
     equation:
   \begin{equation}
    {\cal J}^A_{\mu} \equiv - e \overline{u}_f^A \gamma_{\mu} u_i^A \exp[i(p_f^A-p_i^A)\cdot x_A]
     \end{equation} 
     where $p_i^A$ and $p_f^A$ are the energy-momentum 4-vectors of the incoming
     and scattered electron, respectively, that emit a virtual photon at the space-time
    point $x_A$ and $-e$ is the electron charge. The overall centre-of-mass frame (Fig1b) is a Breit frame for the virtual
    photon, i.e. the latter has vanishing energy:
       \begin{equation} 
      q^{A0} = p_i^{A0}- p_f^{A0} = -  q^{B0} =  p_f^{B0}- p_i^{B0} = 0    
      \end{equation}
      Thus, in this frame, the invariant amplitude may be written:
     \begin{equation} 
       T_{fi} = i \int \frac{{\cal J}^A(x_A) \cdot {\cal J}^B(x_A)}{|\vec{q}|^2} d^4x_A
       \end{equation}
       As shown in the Appendix, use of the Fourier transform:
     \begin{equation} 
        \frac{1}{|\vec{q}|^2} = \frac{1}{4 \pi} \int \frac{d^3 x e^{i\vec{q} \cdot \vec{x}}}{|\vec{x}|}
          \end{equation}
       enables the invariant amplitude to be written as the space-time integral:
   \begin{equation} 
    T_{fi} = \frac{i}{4 \pi} \int dt_A \int d^3 x_A  \int d^3 x_B \frac{{\cal J}^A(\vec{x}_A,t_A) \cdot
     {\cal J}^B(\vec{x}_B,t_A)}{|\vec{x}_B-\vec{x}_A|} 
           \end{equation} 
   It can be seen that the integrand in Eqn(6.6) has exactly the same $j \cdot j/r$ structure as the
    potential energy term in the invariant CEM Lagrangian (2.7). Indeed this is to be expected in the Feynman
    Path Integral (FPI) formulation of quantum mechanics~\cite{Feyn1}.
     The physical meaning of Eqn(6.6) is that
    the total amplitude is given by integration over all spatial positions: $\vec{x}_A(t_A)$,  $\vec{x}_B(t_A)$
    at time $t_A$, and all times $t_A$, of emission and  absorption of a single virtual photon
    \footnote{Thus the simple momentum-space propagator $1/q^2$ of
     Eqn(6.1) is equivalent, in space-time, to the exchange of an infinity of virtual
    photons emitted and absorbed at different spatial positions and times. All these virtual photons however
    have, according to Eqn(6.6), infinite velocity.} in the scattering
    process: $e^-e^- \rightarrow e^-e^-$. Since the virtual photon is not observed, this is just a 
     manifestation of quantum mechanial superposition: a sum of different probability amplitudes with the same 
     initial and final states.
     Notice that the virtual photon propagates with infinite velocity between the spatial positions
     $\vec{x}_A$, $\vec{x}_B$ so that the ambiguity in the direction of propagation of the
     space-like virtual
     photon (see Fig.1b and c) has no relevance. Thus QED predicts that virtual photons 
     produce instantaneous `action at a distance' in the overall centre-of-mass frame of M\o ller scattering.
     This is also implicit in the discussion of CEM in Sections 2 and 3 above, since all forces are
     defined at a fixed time in the CM frame of the interacting charges. The meaning of the
     retarded Li\'{e}nard-Wiechert~\cite{LW} potentials and `causality' in relation to the instantaneous
     forces transmitted by space-like virtual photons is discussed in the concluding section
    of this paper. 
 \begin{figure}[htbp]
\begin{center}
\hspace*{-0.5cm}\mbox{
\epsfysize15.0cm\epsffile{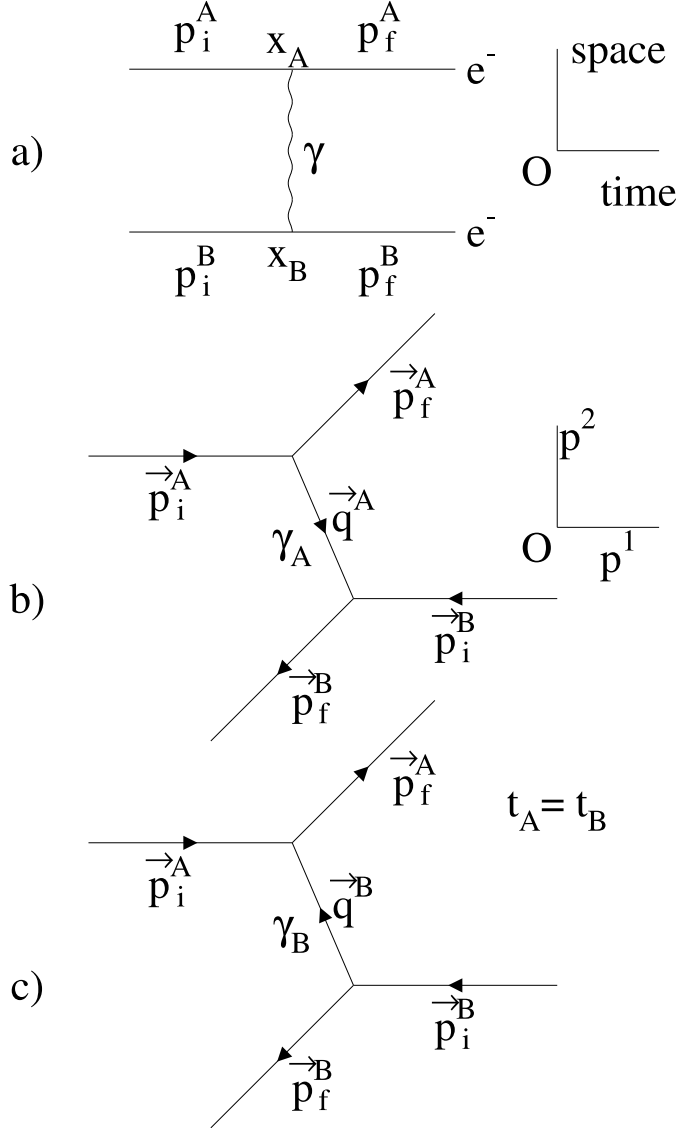}}   
\caption{{\sl a) Feynman diagram for M\o ller scattering: $e^+e^- \rightarrow e^+e^-$, by exchange of a single
  space-like virtual photon. b), c) show the possible momentum space diagrams for M\o ller scattering
  in the CM frame. In b)[c)] the virtual photon transfers momentum from the current ${\cal J}^A$ [${\cal J}^B$]
  to ${\cal J}^B$ [${\cal J}^A$]. These are equivalent descriptions. In both cases the energy of the virtual
  photon vanishes and it has infinite velocity.}}
\label{fig-fig1}
\end{center}
 \end{figure}
    \par To examine more closely the connection between Eqn(6.6) and the FPI formalism, consider the general
       FPI expression for a transition amplitude~\cite{Feyn1}:
      \begin{equation}
      T_{fi}^{FPI} \equiv \langle \chi(t_f)|\psi(t_i) \rangle = \int_{{\rm paths}}
      \chi^{\ast}(x_f,t_f) e^{iS} \psi(x_i,t_i){\cal D} x  \equiv
       \int_{{\rm paths}} \langle f|e^{iS}|i \rangle {\cal D} x   
     \end{equation}
      where the Action, $S$, is given by the time integral of the {\it classical} Lagrangian, $L$, of the
      quantum system under consideration:
      \begin{equation}
       S = \int_{t_i}^{t_f} L(x,\dot{x}) dt \end{equation} 
     (here the upper dot denotes time differentiation) and
      \begin{equation}
      {\cal D} x \equiv {\rm Lim}~(\epsilon \rightarrow 0)~ \frac{d x_0}{A} \frac{d x_1}{A}... \frac{d x_{j-1}}{A}
       \frac{d x_{j}}{A}
     \end{equation}
        where $x_0$,$x_1$,... denote sucessive positions along the path, each separated  by a small, fixed, time
       interval $\epsilon$. Also  $dx_j \equiv x_j - x_{j-1}$. $A$ is a normalistation constant that depends 
       upon $\epsilon$. In the case of present interest, M\o ller scattering, the one dimensional FPI (6.7),
         with a single particle, is generalised to three spatial dimensions and two particles
         with the label $p = A, B$, $x \rightarrow (x_p^1,x_p^2,x_p^3)$, corresponding to the two electrons which
          scatter from each other (see Fig.1).
          In this case, (6.7) is generalised to~\cite{FHPI}:
        \begin{equation}
     T_{fi}^{FPI} =   \int_{{\rm paths}} \langle f|e^{iS}|i \rangle\prod_{p=A,B} \prod_{j = 1}^3 {\cal D} x_p^j(t)  
         \end{equation}
         and (6.8) to
     \begin{equation}
       S = \int_{t_i}^{t_f} L(\vec{x}_A, \dot{\vec{x}}_A;\vec{x}_B, \dot{\vec{x}}_B)dt
      \end{equation}
        where $i$ and $f$ are the initial and final states of the M\o ller scattering process.   
      Assuming that the transition $f \rightarrow i$ is caused by a small term $S_{int}$ in the action where
     $S = S_0+S_{int}$ and $\langle f|S_0| i\rangle = 0$, enables (6.10) to be written as:
     \begin{eqnarray}
      T_{fi}^{FPI} & =  & \int_{{\rm paths}} \langle f|e^{iS_{int}}|i \rangle  \prod_{p=A,B} \prod_{j = 1}^3 {\cal D} x_p^j(t)
          \nonumber \\   
          & = &  \int_{{\rm paths}} \langle f|1 + iS_{int} +\frac{(iS_{int})^2}{2!}+...)
       |i \rangle  \prod_{p=A,B} \prod_{j = 1}^3 {\cal D} x_p^j(t)   \nonumber \\
           & = & i \int_{{\rm paths}}\langle f|S_{int}| i \rangle
          \prod_{p=A,B} \prod_{j = 1}^3 {\cal D} x_p^j(t) + O(S_{int}^2)\nonumber \\
           & \simeq &  i  \int dt \int_{{\rm paths}}\langle f|L_{int}| i \rangle
       \prod_{p=A,B} \prod_{j = 1}^3 {\cal D} x_p^j(t) \nonumber \\
           &\propto &  i  \int dt_A \int d^3 x_A  \int d^3 x_B\langle f| L_{int} | i \rangle 
       \end{eqnarray}
     In the last line the formal differentials ${\cal D} x^i_p(t)$ for arbitary space-time paths
     $x^i_p(t)$ are replaced by those corresponding
     to the electrons A and B in the M\o ller scattering process that, in the classical limit, propagate
    along straight-line paths so that\footnote{Different choices of $\epsilon$ correspond to different values
       of $j$ in Eqn(6.9), for a given value of $\Delta x = x_j -x_0$. In the case of a straight line
       path the value of the limit in (6.9) is independent of the value of $\epsilon$. In particular, the 
       choice $\epsilon = \Delta x/v$ is possible. This yields Eqn(6.13).}:
       \begin{equation}
       \prod_{p=A,B} \prod_{j = 1}^3 {\cal D} x_p^j(t) \propto d^3x_A d^3x_B
    \end{equation} 
     where the normalisation constant in (6.9) has been dropped, since only the proportionality of the
        matrix elements (6.6) and (6.12) is under investigation.
      Comparison of Eqns(6.6) and (6.12) gives:
       \begin{equation}
   \langle f| L_{int} | i \rangle \propto \frac{{\cal J}^A(\vec{x}_A,t_A) \cdot
     {\cal J}^B(\vec{x}_B,t_A)}{4 \pi|\vec{x}_B-\vec{x}_A|}  
    \end{equation}   
    In order to compare Eqn(6.14) with the potential
 energy term in Eqn(2.7) which has the same 4-vector structure as Eqn(6.14), the classical
 limit of the QED transition currents ${\cal J}^A$ and  ${\cal J}^B$, where the momentum carried 
 by the virtual photon vanishes, must be considered. For this it is convenient to use the Gordon
 Identity~\cite{IZ1} for the spinor product appearing in Eqn(6.2):
         \begin{equation}
   \tilde{{\cal J}}^{\mu} \equiv -e \overline{u}_f \gamma^{\mu} u_i =
       \frac{-e}{2m} \overline{u}_f \left[ (p_f + p_i)^{\mu} + i \sigma^{\mu \nu}(p_f- p_i)_{\nu}\right] u_i
        \end{equation} 
    where 
  \[ \sigma^{\mu \nu} \equiv \frac{1}{2}(\gamma^{\mu}\gamma^{\nu}-\gamma^{\nu}\gamma^{\mu}) \]
    In the overall centre-of-mass frame in the limit of vanishing virtual photon momentum:
    $(p_f- p_i)_{\nu} \rightarrow 0$,  $(p_f+ p_i)^{\mu} \rightarrow  2 p^{\mu}$, and
    $ \overline{u}_f u_i  \rightarrow  \overline{u}_i u_i = 2m$~\cite{HM2}. Thus, in this classical
   limit Eqn(6.2) gives:
       \begin{equation}
   \tilde{{\cal J}}^{\mu}_{class} = -e 2 p^{\mu} = 2 Q u^{\mu} = 2 j^{\mu}
         \end{equation}
   where $Q = -e$. Therefore, up to a multiplicative constant, the classical limit of the QED transition 
  current $\tilde{{\cal J}}^{\mu}$ is identical to the CEM current $Q u^{\mu}$ introduced in
   Section 2 above, and also, up to a constant multiplicative factor, the classical limit
  of the matrix element of the QED interaction Lagrangian $\langle f| L_{int} | i \rangle$
   is equal to the
   potential energy term in the CEM Lagrangian (2.7). There is thus a
   seamless transition from QED to CEM. 

  \par Hamilton's Principle of classical mechanics is the $h \rightarrow 0$ limit
   of Feynman's path integral formulation of quantum mechanics. So it may be said that the third postulate in the
   derivation, from first principles, of CEM presented in this paper is not
   really an independent premise, but rather a prediction of quantum mechanics. To show this, it is
   necessary to consider the behaviour of the fundamental FPI formula (6.7) 
  for a transition amplitude in the classical limit. The action S in this formula is a functional of the different
  space-time paths $x(t)$. Writing explicitly the dependence on Planck's constant, gives 
  a multiplicative factor $\exp(iS[x(t)]/\hbar)$ in the transition amplitude. If the paths
   $x(t)$ are chosen such that the variation of $S$ is large in comparison to $\hbar$,
   this factor will exhibit rapid phase oscillations and give a negligible contribution
   to the transition amplitude. If, however, the paths are chosen in such a way that
    $S$ is near to an extremum with respect to their variation, $S$ will change 
    only very slowly from path to path, so that the contributions of different paths
    have have almost the same phase, resulting in a large contribution to the 
    scattering amplitude. The classical limit corresponds to $\hbar \rightarrow 0$,
    where only the path giving the extremum of $S$ contributes. This path is just
    the classical trajectory as defined by Hamilton's Principle. This argument,
    that may be called the `Stationary Phase Principle', was first given by
    Dirac in 1934~\cite{Dirac1}(see also Reference~\cite{Dirac2}) and was an 
    important motivation for Feynman's space-time reformulation of the
    principles of quantum mechanics~\cite{Feyn1,FHPI}.
     \par At  this point it can be truthfully said that there is
    `nothing left to explain' for an understanding of the fundamental physics
    of CEM, given the laws of special relativity  and quantum mechanics.
    The irreducible physical concepts are electrically charged physical objects
    and space-like virtual photons. Coulomb's Law is a consequence of the
    exchange of the latter between the former. Hamilton's Principle is naturally given by
    the classical limit of the FPI formulation of quantum mechanics.

   \par It is interesting, in the light of this `complete understanding' that quantum mechanics 
    and relativity provide about CEM, to consider two further quotations
    from the {\it Principia}. The first is taken from the `General Scholium'~\cite{NP1}.
    After describing the inverse-square law of the gravitational force Newton states:
   \par {\tt But hitherto I have not been able to discover the cause of these \newline properties of gravity
      from phenomena, and I frame no hypothesis, for \newline whatever is not derived from the
     phenomena is to be called a hypothesis, and hypotheses, whether metaphysical 
     or physical, whether of occult qualities or mechanical, have no place in experimental
     philosophy.}
      \par The second is from the `Author's Preface to The Reader'~\cite{NP2}
     \par {\tt I wish we could derive the rest of the phenomena of Nature by the same
       \newline kind of reasoning from mechanical principles, for I am induced by many reasons
       to suspect that they may all depend upon certain forces by which the particles of bodies, by some
      causes hitherto unknown, are either mutually \newline impelled towards one another
      and cohere in regular figures or are repelled and recede from one another.} 
      \par Now, at the beginning of the 21st century, Newton's wish to understand,
       at a deeper level, the forces of nature, has been granted, 
       at least for the case of electromagnetic ones. What was needed was not
      `the same  kind of reasoning from mechanical principles' that Newton considered
     but the discovery of relativity and quantum mechanics. The cause `hitherto unknown' 
      of the electromagnetic force is the exchange of space-like virtual photons according to the
      known laws of QED.
       \par Regrettably science is, at the time of this writing, riddled by many `hypotheses'
      of the type referred to in the first of the above quotations. One such hypothesis, that
      has persisted through much of the 19th century and all of the 20th is that: `No physical
     influence can propagate faster than the speed of light'. This is contradicted by the
     arguments given above and, as discussed in the following section, also by the results of
    some recent experiments. 

  \SECTION{\bf{Discussion and Outlook}}
   The starting point and aims of the present paper are very close to those of Feynman
   and Wheeler when they attempted, in the early 1940's, to reformulate CEM in terms of direct
    inter-charge interactions without the {\it a priori} introduction of any
    electromagnetic field concept. In this way the infinite self-energy terms associated
    with the electric field of a point charge are eliminated. As Feynman put it~\cite{Feyn2}:
     \par{\tt  You see then that my general plan was to first solve the classical \newline problem,
      to get rid of the infinite self-energies in the classical theory, and to hope that when
      I made a quantum theory of it everything would be\newline just fine.}
    \par Feynman and Wheeler had a project to write three papers on the subject~\cite{Feyn3}.
     The first of these papers was to be a study of the classical limit of the 
     quantum theory of radiation.
    Feynman had yet to formulate his space-time version of QED, and this paper
     was never written. In the remaining two papers~\cite{FW1,FW2} it was proposed to introduce
     direct interparticle action by including the effects of both retarded and `advanced' potentials
     as well as an array of `absorbers'. As suggested by Dirac~\cite{Dirac3} half of the difference
     between the retarded
    potential of an accelerated charge and of the `advanced' potential from the absorbers correctly
    predicts the known radiative damping force of CEM. The second paper,~\cite{FW2}, developed
    further this theory by exploiting the Fokker action principle formulation of
     action-at-a-distance in CEM~\cite{Fokker}. As stated in the introduction of
    this paper, a description was being sought that was:
    \begin{itemize}
    \item[(a)] {\tt well defined}
    \item[(b)]  {\tt economical in postulates}
    \item[(c)]  {\tt in agreement with experience}
    \end{itemize}
    that is, in other words, in accordance with Newton's first `Rule of Reasoning in Philosophy' quoted above.
    However, in order to reproduce the known results of CEM by such a theory 
    `advanced' potentials had to be introduced. This immediately gives an apparent breakdown
     of causality and the logical distinction between `past', `present' and `future'.
      As concisely stated by
     Feynman and Wheeler themselves~\cite{FW2}:
     \par {\tt The apparent conflict with causality begins with the thought: if the \newline present motion of
      $a$ is affected by the future motion of $b$, then the \newline observation of $a$ attributes a certain
     inevitabilty to the motion of $b$. Is not this conclusion in conflict with our recognised ability
     to influence \newline the future motion of $b$?}
  \par Feynman and Wheeler then gave a rather artificial example (which the present writer finds unconvincing)
   that was claimed to resolve this causal paradox.
   \par In fact, Feynman and Wheeler were compelled
    to introduce `advanced' potentials because they were assuming, as did also Fokker and earlier
    authors attempting to formulate theories of direct interparticle action in CEM, that causality
    meant that no physical influence could be transmitted faster than the speed of light in
    vacuum. This definition of `causality' seems to have been introduced into physics by
     C.F.Gauss in 1845~\cite{Gauss}. Somewhat later, C.Neumann proposed~\cite{Neumann} that the 
    electric potential responsible for interparticle forces should be transmitted, not
    at the speed of light, but instantaneously, like the gravitational force in Newton's
    theory. As shown above, this is indeed how, in QED, space-like virtual photons 
    transmit the electromagnetic force between charged objects in their common
    CM frame. These two hypotheses will be refered to below below 
    as `Gaussian' and `Neumann' Causality. The fundamental Lagrangian of CEM describing the interaction
    of charged objects, in any  inertial frame, is the simple expression Eqn(2.7) above, not
    the conjectured, and much more complicated,  Fokker action that embodies Gaussian Causality.
      It is important to
    stress that the instantaneous action-at-a-distance, of Neumann Causality, which is just the 
    limit of Gaussian Causality as $c \rightarrow \infty$, unlike an `advanced' potential,
    poses no logical problem of the influence of the future on the present, as
    succinctly stated by Feynman and Wheeler in the above quotation.
    \par Gaussian Causality has been an unstated (and unquestioned) axiom of physics since the advent
   of Special Relativity
   a century ago. The speed
   of light is certainly the limiting velocity of any physical object described
   by a time-like energy-momentum 4-vector. However Einstein at the time when he
   invented special relativity, and Feynman himself, at the time of his collaboration
    with Wheeler, were not aware of the concept of  the `virtual' particles. The latter, 
    associated with the space-time propagators introduced into
  QED by Feynman and Stueckelberg, may be described
   by space-like energy-momentum 4-vectors.  The instantaneous action at a distance of the virtual
  photons in M\o ller scattering described by the invariant amplitude in Eqn(6.6) above,
   can be simply understood from the relativistic kinematics of such virtual particles.
   The relativistic velocity
   $\beta = v/c$ of a particle in terms of its 3-momentum $\vec{p}$ and 4-momentum $p$ is,
   in general, given by the expression:
    \begin{equation}
      \beta = \frac{|\vec{p}|}{\sqrt{\vec{p}^2 + p \cdot p}}
    \end{equation}
    Thus space-like virtual particles, for which, by definition: $ p \cdot p < 0$, are tachyons.
     For the case of the virtual photons exchanged in the center-of-mass-system of M\o ller
    scattering (Figs1b and 1c): $ p \cdot p = -\vec{p}^2$, since $p^0 =0$, and so $\beta$ is infinite,
     consistent with the space-time description in Eqn(6.6).
    \par That the Feynman propagator for a {\it massive} particle 
     violates Gaussian Causality was pointed out by Feynman himself in his first
     QED paper~\cite{Feyn4} and later discussed by him in considerable detail~\cite{Feyn5}.
     This fact is also sometimes mentioned in books on Quantum Field Theory, that otherwise make the
    contradictory claim that, in general, quantum field operators commute for space-like
    separations, so that, in consequence, no physical influence can propagate faster than the
    speed of light\footnote{ For example in Reference~\cite{IZ1} it is stated, in 
    connection with the commutation relation for a pair of scalar fields (Eqn(3.55) of
   ~\cite{IZ1}) that: `Measurements at space time separated points do not interfere
   as a consquence of locality and causality', whereas in the discussion of
   the Feynman propagator $G_F(x)$ in Section 1.3.1 it is stated that: 
     `While the previous Green functions were zero outside the light cone this is not the case for
    $G_F(x)$ which has an exponential tail at negative $x^2$.' The $G_F(x)$ discussed
    here is that corresponding to a classical field , but the corresponding quantum
    propagator, $S_F(x)$, has a similar property~\cite{Feyn4,Feyn5}.}
    The space time propagator of a massive particle was shown by Feynman to be, in 
     general, a Hankel function of the second kind~\cite{Feyn4}. For an on-shell particle, or a virtual
    particle propagating over a large proper time interval, $ \Delta \tau $,
     the propagator has a simpler 
    functional dependence $\simeq \exp(-i m \Delta \tau )$  where $m$ is the pole mass of the particle
    and the proper time interval is defined by the relations:
     \[ \Delta  \tau \equiv \sqrt{\Delta t^2-\Delta x^2}~~~{\rm for}~~~\Delta t^2 \ge \Delta x^2 \]
     \[  \Delta \tau \equiv -i\sqrt{\Delta x^2-\Delta t^2}~~~{\rm for}~~~\Delta x^2 > \Delta t^2 \]
     For space-like separations: $\Delta x^2 > \Delta t^2$ appropriate for the virtual photons
    mediating the Coulomb force, $ \Delta \tau$ is imaginary. This would imply an exponentially 
     damped range of the associated force for the exchange of a massive particle\footnote{For example
      the Yukawa force due to the exchange of virtual pions in nuclear physics.}. Since, however,
     the pole mass of the photon vanishes, no such damping occurs for the exchange of virtual
     photons. The corresponding force law is then the same as for the exchange of real (`on-shell')
     particles, that is, inverse square.
     \par It is instructive to compare Feynman's own discussion of the virtual photon propagator
    in space-time~\cite{Feyn6} to the related one of the invariant amplitude for M\o ller 
    scattering in Section 6 above. Feynman writes out explicitly the 4-vector product in
     Eqn(6.1) to obtain: 
    \begin{equation} 
       T_{fi} =  -i \int \frac{({\cal J}^{A0}{\cal J}^{B0}-{\cal J}^{A1}{\cal J}^{B1}
       -{\cal J}^{A2}{\cal J}^{B2} -{\cal J}^{A3}{\cal J}^{B3})}{q^2} d^4x_A
    \end{equation}   
    Conservation of the current ${\cal J}$ gives the condition:
    \begin{equation}
       q \cdot {\cal J} = q^0 {\cal J}^0-|\vec{q}| {\cal J}^3 = 0
  \end{equation}
   where the 3 axis has been chosen parallel to $\vec{q}$. Use of (7.3) to eliminate $ {\cal J}^{A3}$
   and  $ {\cal J}^{B3}$
  enables (7.2) to be written as:
     \begin{equation} 
       T_{fi} =  i \int\left[ \frac{({\cal J}^{A1}{\cal J}^{B1}
       +{\cal J}^{A2}{\cal J}^{B2})}{q^2}+\frac{{\cal J}^{A0}{\cal J}^{B0}}{\vec{q}^2}\right] d^4x_A
    \end{equation}   
    Feynman then performs a Fourier transform of $(\vec{q})^{-2}$ using Eqn(6.5) to obtain, for the last
    term in the large square  bracket of Eqn(7.4) an equation similar to (6.6) above, but with 
    the replacement: ${\cal J}^{A} \cdot {\cal J}^{B} \rightarrow {\cal J}^{A0}{\cal J}^{B0}$. 
     The instantaneous nature of the Coulomb interaction in this term is noted, but it is also implied
    that the contribution of the transverse polarisation modes:
     $({\cal J}^{A1}{\cal J}^{B1} +{\cal J}^{A2}{\cal J}^{B2})/q^2$ is not instantaneous. Feynman stated:
     \par {\tt The total interaction which includes the interaction of transverse \newline photons then gives
      rise to the retarded interaction.}
     \par This statement is not true when $ T_{fi}$ is evaluated in the CM frame.
      In this case: $q^2 = -\vec{q}^2$, Eqn(6.6) results and the whole interaction of the virtual
      photon is instantaneous. 
\begin{figure}[htbp]
\begin{center}
\hspace*{-0.5cm}\mbox{
\epsfysize15.0cm\epsffile{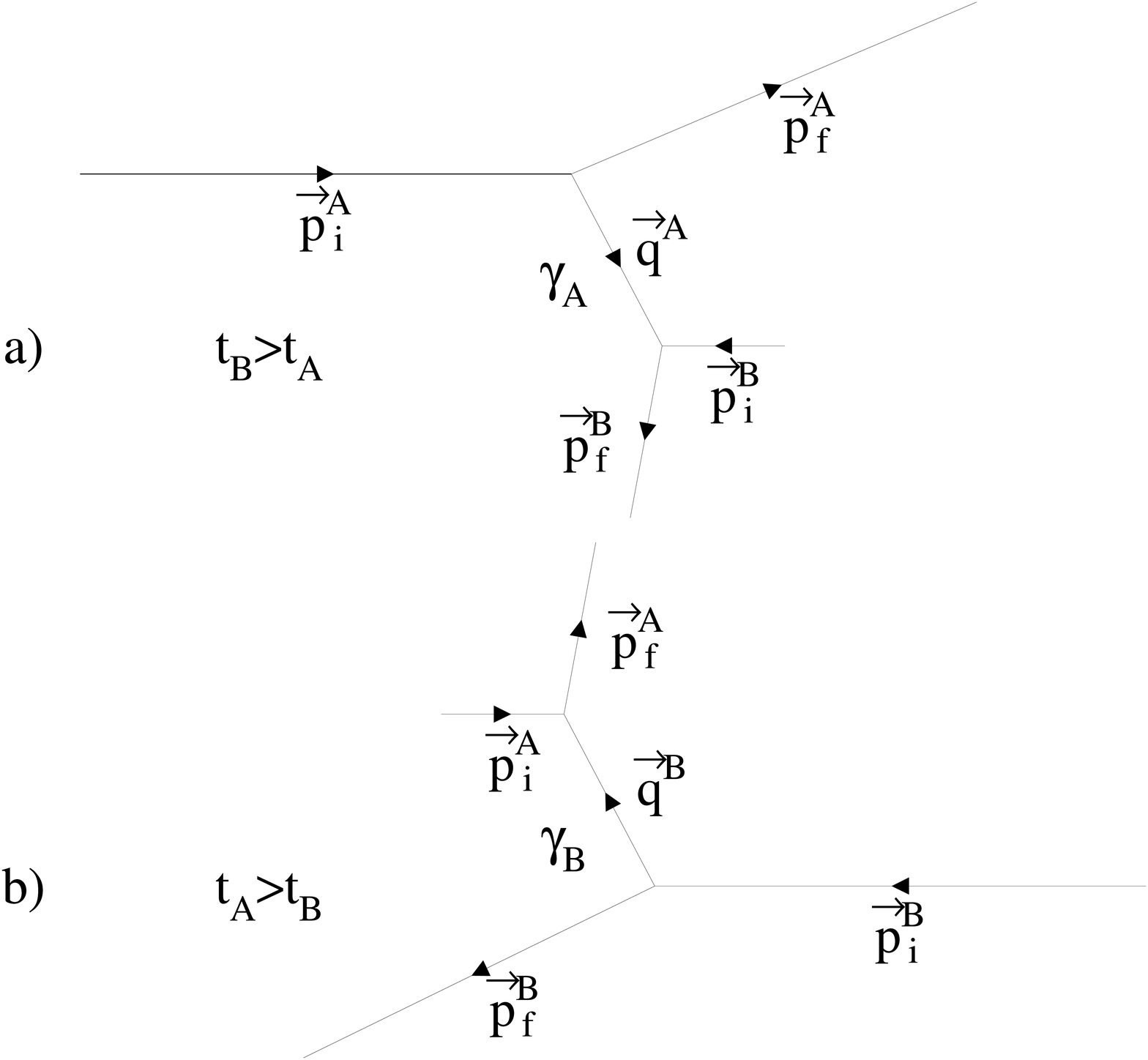}}   
\caption{{\sl Momentum space diagrams for M\o ller scattering of ultra-relativistic electrons
 by $\pi/4$ radians in the CM frame, as in Fig 1 b) and c), as viewed by different
  observers. In a)[b)] the observer is moving parallel to $\vec{p}_i^B$ [$\vec{p}_i^A$]
  with velocity 3c/5 relative to the CM frame. Momentum conservation requires that
  in a) the virtual photon propagates from  ${\cal J}^A$ to ${\cal J}^B$ so that
 $t_B>t_A$, whereas in b) the photon propagates from  ${\cal J}^B$ to ${\cal J}^A$ and
 $t_A>t_B$, where $t_A$ and $t_B$ are the effective times of emission or absorption of the photon
  by the currents ${\cal J}^A$ and ${\cal J}^B$. In both cases the effective velocity $v = pc^2/E$
  of the photon is superluminal: v = 1.044c.}}
\label{fig-fig2}
\end{center}
 \end{figure}
      \par It is amusing to note that a faint `ghost' of Wheeler and Feynman's `advanced' and
      `retarded' potentials subsists in the momentum space diagrams Fig.1b and 1c. 
       The two kinematically distinct situations (i) a virtual photon with momentum $\vec{q}^A$
      propagates from current A to current B (Fig1b) and (ii) a virtual photon with momentum
      $\vec{q}^B = -\vec{q}^A$ propagates from current B to current A (Fig1c) are completely equivalent
      descriptions of the scattering process in the CM frame where $t_A = t_B$. As shown in Fig 2, however,
     this is no longer the the case if the scattering process is observed in a different inertial
     frame.  In Fig 2a the observer is moving with relativistic velocity $\beta = 3/5$ parallel to
     the direction of $\vec{p}_i^B$ in the CM frame. Thus in the observer's proper frame,
      $|\vec{p}_i^B|$ is halved and $|\vec{p}_i^A|$ doubled. In Fig2b, the observer moves with the
     same velocity relative to the CM frame, parallel to $\vec{p}_i^A$. In both cases it follows
    from momentum conservation that there is no possible ambiguity between the momentum space
     configurations shown in Figs 2a and 2b. In Fig 2a the virtual photon must propagate from current A to B
    and so  $t_B > t_A$, and in Fig 2b from current B to A
    so that  $t_A > t_B$. Assuming that the electrons are ultrarelativistic, $E \simeq |\vec{p}|$, and that
    the electrons scatter through $\pi/4$ rad in the CM frame, as shown in Fig 1, the relativistic velocity
    of the virtual photon in the observer's frame is $\beta = 1.044$ for both cases shown in Fig 2. 
    Thus the causal description of scattering processes in momentum space is, in general, frame dependent,
    being ambiguous only in
    the CM frame\footnote{It must not be forgotten, however, that the configurations shown in Fig.2 are 
      in momentum space, not space-time. The different time ordering of `events' in different
     frames that seems apparent on comparing Fig.2a and Fig.2b must therefore be treated
      with caution. In fact, as discussed previously, there is not, in space-time, the exchange
     of single virtual photons with fixed 4-momenta, as seen in Figs 1 and 2, but rather the
     sum over an infinite number of amplitudes corresponding to exchanges of virtual photons
       between all space-time points occupied by the trajectories of the scattered particles,
       as in Eqn(6.6). In the CM frame all such photons have infinite velocity.}. Because of this
      ambiguity, in the kinematical configuration of Fig 1b, the virtual
    photon $\gamma_A$  can be considered as the limit as $t_B-t_A \rightarrow 0$ of an `retarded' interaction
     from A as seen by B, whereas in Fig 1c $\gamma_B$ corresponds to the 
      limit as $t_B-t_A \rightarrow 0$ of an `advanced' interaction produced by B that interacts with A
      \footnote{This corresponds to time increasing from left to right in the momentum space diagrams
      of Fig.1b and 1c, in the same way as in Fig.2a or the Feynman diagram in Fig.1a.}.
     Since the two descriptions are equivalent, the effect is the same as the $t_B-t_A \rightarrow 0$
     limit of half the sum of the retarded interaction produced by the current B and the
     `advanced' interaction produced by the current A. This is the `ghost' of Feynman
     and Wheeler's advanced and retarded potentials mentioned above.
      The current A (B) behaves as the `absorber' for the
     interactions of the current B (A). Unlike in Feynman and Wheeler's formulation however there is no radiation
     and therefore no `radiation resistance'. The photons responsible for the intercharge
     interaction are purely virtual.
     \par As often emphasised by Feynman~\cite{Feyn7}, QED is based on only three elementary amplitudes
     describing, respectively, the propagation of electrons or photons from one space-time point to 
     another and the amplitude for an electron to absorb or emit a photon. The latter is proportional
     to the classical electric charge of the electron. Since only kinematics, and not the coupling
     constant of QED, changes when virtual photons are replaced by real ones it should not be surprising
     if the various field concepts introduced to describe the effect of the virtual photons that generate
     intercharge forces should also be able to provide a description of the observed effects of 
     the creation and absorption of real photons. As will now be shown, this is indeed the case.
     \par A clear distinction should be made however, at the outset, between the fields so far discussed in the present
    paper, representing the effects of virtual photon exchange, and the related fields 
    denoted here as $A_{rad}$, $\vec{E}_{rad}$ and  $\vec{B}_{rad}$ that provide a description of 
    physical systems comprised 
    of large numbers of real photons\footnote{This distinction is usually not
    made in text books on CEM}. As shown in a recent paper by the present author~\cite{JHF3},
    a complex representation of these radiation fields may be identified, in the limit of very low photon
    density, with the quantum wavefunction of a single real photon\footnote{This wavefunction occurs for example,
    in the construction of invariant amplitudes of all processes in which real photons are
    created or destroyed. The related problem of 'non localisability' of photons is also discussed
    in Reference~\cite{JHF3}.}
    \par The electrodynamic Maxwell equation (4.20) as written above therefore describes only
    the effects of virtual photon exchange. All fields and currents are defined at some unique
    time in the CM frame of the interacting charges. The solutions of this equation, $\vec{E}$,~$\vec{B}$  are given
    by Eqns(3.11), (3.12) respectively and (3.1). To arrive at a description of real photons it is 
   convenient to express the electrodynamic Amp\`{e}re Law of Eqn(4.20) uniquely in terms of the 3-vector potential
    by using the Lorenz Condition (4.12) to eliminate the scalar potential $A^0$.
     The result of this simple exercise in 3-vector algebra, which may be found in
    any text-book on CEM, is:
      \begin{equation}
      -\nabla^2\vec{A}_{rad} +\frac{1}{c^2} \frac{\partial^2 \vec{A}_{rad}}{\partial t^2}
      = 4 \pi \vec{j}_{rad}
      \end{equation}
       The `radiation' suffix has been added to $\vec{A}$ and $\vec{j}$ to distinguish them 
       from the quantities  $\vec{A}$ and $\vec{j}$ defined in Eqns(3.1) and (4.8) since the 
    latter are not solutions of Eqn(7.5) unless $c$ is infinite.
    Similarly by using the Lorenz condition to eliminate $ \vec{A}$ in favour
    of $A^0$ the inhomogeneous D'Alembert  equation  for the scalar potential may be 
   derived:
      \begin{equation}
      -\nabla^2 A^0_{rad} +\frac{1}{c^2} \frac{\partial^2  A^0_{rad}}{\partial t^2}
      = 4 \pi j^0_{rad}
      \end{equation}  
    As shown for example in Reference~\cite{Jackson}, the solutions of Eqn(7.5) and (7.6) are similar to (3.1)
     except that they are retarded in time:
      \begin{equation}
      \vec{A}_{rad}(t) = \left\{\frac{\vec{j}}{c(r-\frac{\vec{v} \cdot \vec{r}}{c})}\right\}
       _{t-\frac{r}{c}}
       \end{equation}
      \begin{equation}
      A^0_{rad}(t) = \left\{\frac{j^0}{c(r-\frac{\vec{v} \cdot \vec{r}}{c})}\right\}
       _{t-\frac{r}{c}}
       \end{equation}
   where the large curly bracket indicates that $\vec{j}$ and $r$ are evaluated at the retarded time
   $t-r/c$. It follows that $ \vec{A}_{rad}(t)$ and the associated 
     electromagnetic fields  $\vec{E}_{rad}(t)$ and  $\vec{B}_{rad}(t)$  describe some physical
    effect produced by the source current at time  $t-r/c$, i.e. that propagates from the
    source to the point of observation with velocity $c$. In reality, the energy-momentum
   flux, associated with the corresponding `electromagnetic wave' produced by the source,
   consists of a very large number of real photons whose energy distribution depends on the
   acceleration of the source at their moment of emission. Thus the solutions 
  (7.7) and (7.8) imply the existence of massless physical objects (`photons')~\cite{LLB,JHF4},
   created by the source current. As discussed in Reference~\cite{JHF3}, comparison of the 
   known properties of both photons and the classical electromagnetic waves associated with the
   fields $A_{rad}$, $\vec{E}_{rad}$ and  $\vec{B}_{rad}$ enables many fundamental concepts
   of quantum mechanics to be understood in a simple way. 

   \par Text books and papers on CEM do not usually make the above distinction between the fields
   $\vec{E}$ and $\vec{B}$, describing the mechanical forces acting on charges, and 
   $\vec{E}_{rad}$ and  $\vec{B}_{rad}$ that provide the classical description of radiation
   phenomena, employing identical symbols for both types of fields. An important exception to this 
   is the work of Reference~\cite{CSR}. In this paper,
   the instantaneous nature of the interactions mediated by the  $\vec{E}$ and $\vec{B}$ fields,
   derived in the previous section from QED, is conjectured. These fields are solutions of
   the Maxwell equations: (4.4), (4.6), (4.7) and (4.20). Different, retarded, fields,
   solutions of the D'Alembert equation, equivalent to  $\vec{E}_{rad}$ and  $\vec{B}_{rad}$,
    denoted as  $\vec{E^*}$ and $\vec{B^*}$ were also introduced. The application of the 
    Poynting vector and spatial energy density formulae uniquely to the fields
    $\vec{E^*}$ and $\vec{B^*}$ was pointed out. However, instead of the formulae 
    (7.7) and (7.8) above, only `sourceless' solutions of the homogeeous  D'Alembert equation
     were considered. Also it was proposed, instead of the formulae (3.15) and (3.16) above,
   to define $\vec{E}$ and $\vec{B}$ as the standard `present time'  Li\'{e}nard and Wichert
   formulae\footnote{ See, for example, Reference~\cite{PP}.} which, for a uniformly moving charge,
   are actually equivalent to {\it retarded} fields. The discussion of Reference~\cite{CSR}.
   was carried out entirely at the level of classical fields, considered as solutions of
   partial differential equations with certain boundary conditions. No identification of
   $\vec{E}$ and $\vec{B}$ with the exchange of virtual photons and  $\vec{E^*}$ and $\vec{B^*}$
   as the classical description of real photons was made. The suggestion that $\vec{E}$ and $\vec{B}$
   should be associated with exchange of virtual photons `not subject to causal limitations'
    has, however, been made in a recent paper~\cite{ALK}

    \par The electric and magnetic fields derived from the  Li\'{e}nard and Wiechert potentials
    (7.7) and (7.8) contain terms with both $1/r^2$ and $1/r$ dependencies. Both fields are
    retarded, but conventionally only the latter are associated with radiative effects (the fields 
      $\vec{E}_{rad}$ and  $\vec{B}_{rad}$) in CEM. It is interesting to note that there
    is now mounting experimental evidence~\cite{Walker,KMSTCM}, that the fields $\simeq~1/r^2$
     are instantaneous and not retarded, and so should be associated with the force fields
      $\vec{E}$ and $\vec{B}$ mediated by virtual photon exchange. Particularly convincing are the
    results shown in Reference~\cite{KMSTCM} where the temporal dependence of near- and far-magnetic
     fields were investigated by measuring electromagnetic induction at different distances from
    a circular antenna. Figure 8 of~\cite{KMSTCM}apparently shows clear evidence for the instantaneous
    nature of the $1/r^2$ `bound fields' (i.e. fields associated with virtual photon exchange). This suggests
    that the retarded  $1/r^2$ solutions of (7.5) and (7.6) should be discarded as unphysical, 
      whereas the retarded  $1/r$ solutions describing correctly the `far-field' in the
      experiment~\cite{KMSTCM} do give the correct classical description of the radiation of real
    photons. There seems now to be therefore experimental evidence for electromagnetic fields
    respecting both
    Neumann causality (the force fields $\vec{E}$ and $\vec{B}$) as well as Gaussian causality
    (the radiation fields $\vec{E}_{rad}$ and  $\vec{B}_{rad}$). 

   \par Maxwell's original discovery of electromagnetic waves~\cite{Maxwell} was based on an equation 
   similar to (7.5) for components of the electromagnetic fields, but without any source term, which
   is just the well-known classical Wave Equation in three spatial dimensions
    Although this procedure leads, in a heuristic manner, to the concept of
    `electromagnetic waves' propagating at speed $c$, with vast practical, political and
    sociological consequences, it can be seen, with hindsight, to have been a mistake from 
   the viewpoint of fundamental physics. In fact, if the current vanishes, so, by definition,
   do all the fields whether instantaneous as in Eqn(3.1) or retarded as in Eqn(7.5). If all the
    fields vanish there can evidently be no `waves'.
    The result of this mistake was many decades of fruitless work by Maxwell and others
    to invent a medium (the luminiferous aether) in which such `sourceless' waves might
   propagate and whose properties would predict the value of $c$. Now it is understood that
   the energy density $(\vec{E}_{rad}^2+\vec{B}_{rad}^2)/8\pi$ of a plane `electromagnetic wave'
   is simply that of the beam of real photons of which it actually consists~\cite{JHF3}.
    \par The existence of photons, massless particles with constant velocity c, is predicted
   by Eqn(7.5) that necessarily follows from Eqns(4.12) and (4.20). These in turn may be
   derived from the Lagrangian (2.7) and Hamilton's Principle. It is then interesting to ask
   where the constant `c' was introduced into the derivation. The answer is Eqn(2.2), the definition
   of 4-vector velocity.  The same formula contains, implicitly, the information that a massless 
   particle has the constant velocity, c , that is used to identify the 'electromagnetic wave',
   with velocity c predicted by Eqns(7.5) and (7.6), with the propagation of the massless real photons
   produced by the source.
   \par The only dynamical assumption in the derivation of CEM presented above is Coulomb's Law.
    If it is explained in QED as an effect due to virtual photon exchange, it also seems to require
    via Eqns(7.5) and (7.6), the existence of real, massless, photons. Although clearly of interest,
    the further study of the relationship between CEM and QED for radiative processes is, as stated
   earlier, beyond the scope of the present paper.
   \par In conclusion, the results obtained in the present paper are compared with those of the 
   similarly motivated project of Feynman and Wheeler. The latter made the following general comments
   on their approach~\cite{FW2}:
    \par {\tt (1)~ There is no such concept as ``the'' field, an independent entity
           \newline with degrees of freedom
      of its own. }
     \par {\tt (2)~There is no action of an elementary charge upon itself and
     \newline consequently no problem of
     an infinity in the energy of the electromagnetic\newline field.}
    \par {\tt (3)~The symmetry between past and future in the prescription of the
     \newline fields not a mere
     logical possibility, as in the usual theory, but a
     \newline postulational requirement.}
     \par The statements (1) and (2) remain true in the approach described in Sections 2-4 above. However
    the writer' opinion is that the `infinite self energy' problem of CEM is really an artifact
    of the possibly unphysical concept of a `point charge' rather than a shortcoming of the 
   classical electromagnetic field concept {\it per se}. That being said, it remains true that
   the virtual photons interacting with a given charge are produced by other charges so there
   is no way for the charge to `interact with itself'. If the energy of the `electromagentic field'
    is identified with that of the exchanged virtual photons in the CM frame, it vanishes, so, there
   is, as in (2) above, certainly no self energy problem. However, the statements (1)and (2) are only
   applicable to the `force' fields introduced in Eqns(3.1), (3.10) and (3.12) above, that may be 
   denoted as $A_{for}$, $\vec{E}_{for}$ and  $\vec{B}_{for}$ to distingish them from the 
   `radiation' fields describing real photons. It is important to reiterate that the
    definitions and physical meanings of these two types of fields are quite distinct. The 
    quantity: $(\vec{E}_{for}^2+\vec{B}_{for}^2)/8\pi$ does not correctly describe the energy
   density of the electromagnetic field associated with virtual photons, and, in contradiction to (1),
   extra degrees of freedom must be added to the Lagrangian to correctly describe real photons.
    No distinction was made between real and virtual photons by  Feynman and Wheeler.
   In the approach of the present paper, point (3) with its introduction of acausal `advanced'
   potentials is no longer valid. It was a consequence of Feynman and Wheeler's taking Gaussian
   Causality as an axiom. The latter is true, as shown by Eqn(7.7) and (7.8), for any interaction
   transmitted by real photons (i.e. for the fields  $A_{rad}$, $\vec{E}_{rad}$ and  $\vec{B}_{rad}$)
   but not, as shown in Section 5 above, for the force fields describing the effects of
   the exchange of space-like virtual photons. These are always tachyonic (as in Fig.2) and
   may be instantaneous (as in Fig1b and c) but do not, unlike `advanced potentials',
   violate causality. Feynman and Wheeler's mistake, the same as that of many
    previous authors, was to try to describe the physical
   effects of virtual photon exchange by fields respecting Gaussian, instead of Neumann, Causality.
    \par It is instructive to compare the discussion of CEM in the present paper with that
    of Reference~\cite{LL}, which also takes as fundamental physical assumptions, in constructing
    the theory, special relativity and Hamilton's Principle. However in Reference~\cite{LL},
    the existence of the 4-vector potental $A$ and the relativistic Lagrangian equivalent 
    to (3.2) above are both
    postulated {\it a priori}. This procedure is justified by the statement~\cite{LL1}:
 \par {\tt The assertions which follow should be regarded as being, to a certain \newline extent,
    the  consequence of experimental data. The form of the action for \newline a particle in an
   electromagnetic field cannot be fixed on the basis of \newline general considerations alone 
   (such as, for example the requirement of \newline relativistic invariance). }
   \par This is true, as far as it goes, but fails to take account of either the constructive
   principle put forward in the quotation from Hagedorn cited above, or the known essential physics 
   of the problem embodied in the inverse-square force law between charges in the static limit.
    As demonstrated in Section 2 above, the assumption of this law, together with the classical 
   definition of potential energy and relativistic invariance is in fact sufficient to derive
   just the Lagrangian that is assumed {\it a priori} in~\cite{LL}. The derivations of the
    Lorentz force equation and the covariant definitions of electric and magnetic fields
    (3.11) and (3.12) given in~\cite{LL} are identical to those presented above, as are also the
    derivations of the magnetostatic Maxwell equation and the Faraday-Lenz law. In Chapter 3 of
   ~\cite{LL} there is a lengthy discussion of the motion of particles in magnetic fields.  However
   at this point the magnetic field is a purely abstract mathematical concept. How it may be obtained
    from its sources $-$ charges in motion $-$  has still not been even mentioned! Only after the
    electrostatic and electrodynamic Maxwell equations have been derived in Chapter 4 from the principle
    of least action, by treating the electromagnetic fields as `co-ordinates', is the relation between
    fields and
    their sources established. Coulomb's law is then derived at the begining of Chapter 5 (page 100!)
    from the Poisson equation. In contrast, in the present paper, Coulomb's law (and hence the
    Poisson equation) is assumed at the outset, and the electromagnetic Maxwell equation is
    derived, simply by inspection, from the covariant form of Poisson's equation. At this point identical
   results have been obtained from the same essential input (the Lagrangian (3.2)) by the present
   paper and~\cite{LL}. However the present writer feels that there are enormous pedagogical advantages,
    (especially in view of the crucial role of Coulomb's law in QED, discussed above) to start the
    discussion with the vital experimental fact $-$ the inverse-square force law $-$ rather than
    to derive it after 100 pages of complicated mathematics, as is done in~\cite{LL}. Also, in~\cite{LL}
    no distinction is made between  $A_{for}$, $\vec{E}_{for}$, $\vec{B}_{for}$ and $A_{rad}$, $\vec{E}_{rad}$,
    $\vec{B}_{rad}$. All fields
    are assumed to be derived from the same, non-relativistic, retarded,  Li\'{e}nard and Wichert potentials.
    \par Finally the approach of the present paper may be compared with that of another recent paper by
    the present author~\cite{JHF1} in which the Lorentz Force Law, magnetic field concept and
    the Faraday-Lenz Law are
    derived from a different set of postulates. The electrostatic definition of the electric
    field $\vec{E}_{stat} = -\vec{\nabla} V$ is first generalised to the covariant form of Eqn(3.11)
    above by imposing space-time exchange symmetry invariance~\cite{JHF2}. The magnetic field concept
   and the Lorentz Force Law are then shown to follow from the covariance of Eqn(3.11), and the
   derivation of the Faraday-Lenz law is identical to that given above.
   Neither Coulomb's Law nor Hamilton's Principle were invoked in this case, demonstrating the robustness
   of some essential formulae of CEM to the choice of axioms for their derivation. Another example of this
   is provided by Reference~\cite{LL} where Coulomb's law is derived from the principle of least 
   action and the relativistic Lagrangian (3.2), as initial postulates.
   \par{ \bf Acknowledgement}
   \par I thank B.Echenard and P.Enders for their comments on this paper. Pertinent and constructive
     critical comments by an anonymous referee have enabled me to simplify, or improve, the presentation
      in several places. They are gratefully acknowledged.
  \newpage
{\bf Appendix}
 \par Factoring out the space-time dependent factor in the transition current ${\cal J}_{\mu}$ according
  to the definition
       \begin{eqnarray}
   {\cal J}_{\mu} & = & -e \overline{u} \gamma_{\mu} u \exp[i(p_f-p_i) \cdot x]  
   \nonumber \\
   & \equiv & \tilde{{\cal J}}_{\mu}  \exp[i(p_f-p_i) \cdot x]~~~~~~~(A1) \nonumber
        \end{eqnarray}
   enables the invariant amplitude $T_{fi}$ of Eqn(6.4) to be written as:
      \begin{eqnarray}
      T_{fi} & = & i \int dt_A \int d^3x_A \frac{\tilde{{\cal J}}^A e^{i(p_f^A-p_i^A) \cdot x_A}\cdot
       \tilde{{\cal J}}^B e^{i(p_f^B-p_i^B) \cdot x_A}}{|\vec{q}^2|}    \nonumber \\              
      & = & i \int dt_A \int d^3 x_A \frac{\tilde{{\cal J}}^A e^{i(p_f^{A0}-p_i^{A0})t_A}\cdot
       \tilde{{\cal J}}^B e^{i(p_f^{B0}-p_i^{B0}) t_A}}
     {|\vec{q}^2|}~~~~~~~(A2)  \nonumber
     \end{eqnarray}
       since, from momentum conservation:
      \[ \vec{p}_f^A-\vec{p}_i^A = - (\vec{p}_f^B-\vec{p}_i^B)~~~~~~~(A3) \]
    Using now Eqn(6.5) gives 
 \[   T_{fi}  =  i \int dt_A \int d^3x_A 
  \tilde{{\cal J}}^A e^{i(p_f^{A0}-p_i^{A0})t_A}\cdot
       \tilde{{\cal J}}^B e^{i(p_f^{B0}-p_i^{B0})t_A}
      \int \frac{e^{ i\vec{q} \cdot \vec{x}}d^3x}{4 \pi |\vec{x}|}~~~~~~~~(A4)  \]     
      Making the change of variables:
    \[ \vec{x} = \vec{x}_B- \vec{x}_A,~~~d^3x = d^3x_B \]
  and noting that $\vec{q} = \vec{p}_i^A- \vec{p}_f^A$ gives, from Eqn(A4):
   \[  T_{fi}  =  i \int dt_A \int d^3x_A
  \tilde{{\cal J}}^A e^{i(p_f^{A0}-p_i^{A0})t_A}\cdot
       \tilde{{\cal J}}^B e^{i(p_f^{B0}-p_i^{B0})t_A} 
      \int \frac{e^{ i(\vec{p}_f^A- \vec{p}_i^A) \cdot (\vec{x}_B -\vec{x}_A) }d^3x_B}
    {4 \pi |\vec{x}_B- \vec{x}_A|}~~~~~~~~(A5)  \]        
    Now 
   \[(\vec{p}_f^A- \vec{p}_i^A) \cdot (\vec{x}_B -\vec{x}_A)  =   -(\vec{p}_f^B- \vec{p}_i^B) \cdot \vec{x}_B
     -(\vec{p}_f^A- \vec{p}_i^A) \cdot \vec{x}_A~~~~~~~~(A6) \]
  where Eqn(A3) has been used. 
     Substituting (A6) into (A5), yields Eqn(6.6) of
     the text.    
\pagebreak
 
\end{document}